\documentclass[sigconf,9pt]{acmart}

\usepackage[ruled,linesnumbered]{algorithm2e}

\usepackage{CJKutf8}  

\usepackage{threeparttable}
\usepackage{pifont} % used for \ding{xx}

\newcommand{\cmark}{\textcolor{green!80!black}{\ding{51}}}
\newcommand{\xmark}{\textcolor{red}{\ding{55}}}

\usepackage{xcolor}
\usepackage{pgfplots}
\usepackage{tikz}
\usepackage{comment}

\usepackage{filecontents}

\usepackage{url}
\usepackage{hyperref}

\usepackage{multirow}
\usepackage{amsmath,amsfonts}

\usepackage{textcomp}
\usepackage[most]{tcolorbox}
\definecolor{ao(english)}{rgb}{0.0, 0.5, 0.0}

\usepackage{subfigure}
\usepackage{filecontents}

\makeatletter
\renewcommand\@formatdoi[1]{\ignorespaces}
\makeatother

\usepackage{graphicx,subfigure}

\definecolor{bblue}{HTML}{4F81BD}
\definecolor{rred}{HTML}{C0504D}
\definecolor{ggreen}{HTML}{9BBB59}
\definecolor{pviolet}{HTML}{9F4C7C}

\usepackage{listings}

\usepackage{blindtext}
\usepackage{etoolbox,xstring,mfirstuc,textcase}
\usepackage{booktabs}
\usepackage{pgfplots}

\usepackage{tabularx,booktabs,makecell,multirow, caption}
\usepackage{enumitem} % to control Itemization spacing

\newcolumntype{L}{>{\arraybackslash}X}

\usepackage{tikz}

\definecolor{findOptimalPartition}{HTML}{D7191C}
\definecolor{storeClusterComponent}{HTML}{FDAE61}
\definecolor{dbscan}{HTML}{ABDDA4}
\definecolor{constructCluster}{HTML}{2B83BA}

\renewcommand\footnotetextcopyrightpermission[1]{} % removes footnote with conference information in first column

\acmConference[IMC'23]{ACM Internet Measurement Conference}{Oct 24-26}{Montreal, CA}
\begin{document}
%=================================================
\title{An Empirical Study on Snapshot DAOs}
%=================================================   

\author{Qin Wang$^{1}$, Guangsheng Yu$^{1}$, Yilin Sai$^{1}$, Caijun Sun$^{2}$,  \\Lam Duc Nguyen$^{1}$, Sherry Xu$^{1}$, Shiping Chen$^{1}$}

\affiliation{
\textit{$^1$CSIRO Data61, Australia} \\
\textit{$^2$Zhejiang Lab, China} \\
}

\begin{abstract}
The notion of Decentralized Autonomous Organization (DAO) is an organization constructed by automatically executed rules such as via smart contracts, holding features of the permissionless committee, transparent proposals, and fair contributions by stakeholders. As of May 2023, DAO has impacted over \$24.3B market caps. However, there are no substantial studies focused on this emerging field. To fill the gap, we start from the ground truth by empirically studying the breadth and depth of the DAO markets in mainstream public chain ecosystems in this paper. We dive into the most widely adoptable DAO launchpad, \textit{Snapshot}, which covers 95\% in the wild DAO projects for data collection and analysis. By integrating extensively enrolled DAOs and corresponding data measurements, we explore statistical resources from Snapshot and analyze data from 581 DAO projects, encompassing 16,246 proposals over the course of 5 years. Our empirical research has uncovered a multitude of previously unknown facts about DAOs, spanning topics such as their status, features, performance, threats, and ways of improvement. We have distilled these findings into a series of key insights and takeaway messages, emphasizing their significance. Notably, our study is the first of its kind to comprehensively examine the DAO ecosystem.

\end{abstract}

%\begin{IEEEkeywords}
%\keywords{Blockchain, DAO, Snapshot}
%\end{IEEEkeywords}

%=================================================
\maketitle
%=================================================  

%-----------------------------------------
\section{Introduction}
%-----------------------------------------

Decentralized autonomous organizations (DAOs) emerge with the rapid development of cryptocurrency and blockchain. DAO is an entity that is collaboratively managed by on-chain participants to deploy resources, release proposals and make decisions. The usage of DAO in governance can decentralize the operation via blockchain by enabling on-chain rules and activities transparent and traceable. Every stakeholder is eligible to propose, vote, and enact changes for DAO proposals. As of May 2023, a total of 12,824 organizations have been created. The invested funding towards these DAOs (a.k.a., treasury) has reached up to \$24.3B, while engaged members increased 531x from 13K to 6.9M members during the last 6 years\footnote{Data source: \url{https://deepdao.io/organizations}. [May 2023]}. Among them, 4.5M participants are active voters or proposal makers. DAOs accordingly become a force to be reckoned with in the Web3 space~\cite{wang2022exploring} and new cryptocurrency markets.

\smallskip
\noindent\textbf{The open problem.} Although DAOs have gained traction in recent years, their structural development is still in its nascent stage. One of the primary challenges is that DAO projects are diverse, with varying objectives and functionalities. Some DAOs begin with a clear purpose, such as Uniswap and Bancor, which can remain focused on serving a specific community or users. In contrast, other DAOs may diverge from their initial objectives. This means a DAO may begin with a simple goal like collecting NFTs, and then morph into a community to attract participants (e.g., PleasrDAO), a trading platform to trade NFTs (Opensea), or an incubator to invest artists (BAYC). The variety of forms and outcomes can be confusing for newcomers and experts alike in the blockchain space. Creating a comprehensive and structural view of DAOs is still a significant challenge that requires further development and refinement.

Recent studies have attempted to examine DAOs (cf. \underline{Table~\ref{tab-comp}}). Several studies (see Line~1) begin by providing an overview of existing literature. However, due to the delay in publication, academic works may lack current and persuasive examples of DAO projects, leading to outdated information (equiv. \textit{\textbf{in-time}} or not). Other works  (Line~2) propose a high-level framework to discuss DAO properties, but their abstracted metrics are created before events, making them impractical (close-to-real or \textit{\textbf{applicable}}?). Several studies (Line~3) focus on capturing features from real DAO projects through empirical research, but their sample pools are often limited (resourceful or investigated projects in large \textit{\textbf{scale}}?). All of these efforts seem to fall short of providing a comprehensive and up-to-date understanding of DAOs for readers.

\begin{table}[!hbt]
\vspace{-0.12in}
\centering
\caption{This work \textcolor{gray}{v.s.} DAO Studies}\label{tab-comp}
\vspace{-0.15in}
\resizebox{\linewidth}{!}{
\begin{tabular}{cccccc}
\toprule
\multicolumn{1}{c}{\textit{\textbf{Examples}}}& \multicolumn{1}{c}{\textit{\textbf{Method}}} &
\multicolumn{1}{c}{\textit{\textbf{Target}}} &\multicolumn{1}{c}{\textit{\textbf{In-time}}} & \multicolumn{1}{c}{\textit{\textbf{Applicable}}} & \multicolumn{1}{c}{\textit{\textbf{Scale}}} \\ 
\midrule

\cite{liu2021technology}\cite{yu2022leveraging}\cite{faqir2021comparative}\cite{wang2019decentralized} & \multirow{1}{*}{Literature review} & Publications & \textcolor{red}{Not very} & n/a & \textcolor{red}{<30} \\

\cmidrule{2-3}

\cite{bischof2022longevity}\cite{mehdi2022data} & \multirow{1}{*}{Framework} & Properties & n/a & \textcolor{red}{Priori} & n/a   \\

\cmidrule{2-3}

\cite{feichtinger2023hidden}\cite{fritsch2022analyzing}\cite{sharma2023unpacking} & \multirow{1}{*}{Empirical study} & Projects & \textcolor{blue}{In-time} & \textcolor{blue}{Practical}  & \textcolor{red}{<22} \\

\cmidrule{2-3}

\textit{\textbf{This work}}  & Empirical study & \textcolor{blue}{Launchpad} & \textcolor{blue}{In-time} & \textcolor{blue}{Practical} & \textcolor{blue}{>500}  \\
\bottomrule
\end{tabular}
}
\vspace{-0.13in}
\end{table}

\smallskip
\noindent\textbf{Our attempts.} To address the aforementioned shortcomings, we have devised a unique approach to our study. After conducting thorough research, we have found that existing DAO launchpads and DataFeeds have amassed a wealth of information on numerous DAO projects, including both long-standing DAOs that have ceased operations and newly-launched ones that have appeared within the last month. In an effort to avoid duplicating the work of others, we have opted to omit DataFeeds and some launchpads that have already presented analyzed data. Instead, we are focusing our attention on a lesser-known launchpad, namely \textit{Snapshot}~\cite{snapshot}, which has compiled a significant number of DAOs but has yet to conduct extensive analyses on them.

The emergence of Snapshot is to overcome the issue of high costs associated with on-chain operations due to the complexity of consensus and frequent voter interactions. Snapshot accordingly introduced an off-chain voting tool that enables practitioners to efficiently access popular DAOs for voting, managing, auditing, and researching. Snapshot serves as a launchpad that captures over 95\% of in-the-wild DAO projects (over 11,000 spaces)  and offers open access to create new DAOs that are compatible with mainstream blockchain platforms like Ethereum ~\cite{wood2014ethereum}, Avalanche \cite{avalanche}, Binance smart chain (BSC)~\cite{bsc}, Polygon~\cite{polygon}, Solana~\cite{solana}, and more. The ample and reliable data collected by Snapshot motivate us to develop the following in-depth as well as comprehensive research surrounding DAOs.

\smallskip
\noindent\textbf{Contributions.} In this paper, we dive into the DAO projects that are created and managed on Snapshot. We develop the research by gradually approaching the DAO basic concept, operating mechanism, and relevant techniques, and analyzing the statistical data collected from Snapshot. Our work is the first study to strictly explore the features of DAOs, providing in-time guidance for the following readers. Specifically, we detail our contributions here. 

\noindent\hangindent 1em\textit{$\triangleright$ A structural investigation on DAOs} (\textcolor{magenta}{Sec.\ref{sec-dao}}). Intending to be a complete study focused on DAOs, we clear the fog surrounding this fuzzy term by presenting its underlying structures (e.g., components, supportive standards), core mechanisms, and outstanding instances. In particular, based on extensive investigation, we decouple DAO constructions (e.g., decentralized identifier, utility token, smart contract, e-voting) and extract a series of metrics to reflect the features (details refer to \textcolor{magenta}{Sec.\ref{subsec-framework}}) in DAO's designs. As tokenization plays an essential role in DAO governance (discussed in \textcolor{magenta}{Sec.\ref{subsec-tokenization}}), we also sort out the relevant token standards that are essential to the DAO's incentive. Further, we provide a short list of tools (cf. \underline{Table~\ref{tab-summary}}) that well support DAO operations. 

\noindent\hangindent 1em\textit{$\triangleright$ A comprehensive exploration on Snapshot launchpad} (\textcolor{magenta}{Sec.\ref{sec-snapshot}}). We study one mainstream DAO-related governance tool, Snapshot, by summarising the features of \textit{involved entities}, \textit{running mechanisms}, \textit{typical operations}, and \textit{voting strategies}. As of Nov. 2022\footnote{A notable distinction between two dates: Nov. 2022 marks the conclusion of the experiments, while May 2023 represents the nearing completion of writing the paper.}, Snapshot has registered 11K+ spaces (projects), which covers 95\% \textit{in-the-wild} DAOs. However, many of them are inactive with very few members or proposals. We ignore such projects and put our focus on the influential ones. Thus, we collect the 581 most prevalent DAO projects that contain a total of 16,246 proposals over the span of the past 5 years. In particular, we dive into each project and scrutinize included proposals with basic information, voting strategy, proposal content, and voting results.

\noindent\hangindent 1em\textit{$\triangleright$ A solid analysis for collected data from Snapshot} (\textcolor{magenta}{Sec.\ref{sec-experi}}). Based on extensive investigation and exploration, we structure our experimental results from four aspects that separately interpret the project scale, supporting infrastructure, dependent e-voting schemes, and the operational tokens (cf. \underline{Table~\ref{tab-result}}). We evaluate each item by diving into multiple sub-aspects. Accordingly, we study the details of DAO members (e.g., number of participants), basic project information (project duration, language usage, storage condition, underlying platform), and voting process (voting pattern, results, distribution, variances, token usage, meaningful contexts). With substantive evidence, we conclude seven pieces of \textit{home-taking} messages (\textbf{\text{Insights \ding{202}-\ding{208}}} in grey banners and the tail of last page) as high-level summaries for interested readers.

\noindent\hangindent 1em\textit{$\triangleright$ A series of reasonable analyses and discussions for building better DAOs} (\textcolor{magenta}{Sec.\ref{sec-discuss}}\&\textcolor{magenta}{\ref{sec-finding}}). Existing DAO fields are absent of rigorous studies that can deliver effective educational guidance. We thus provide our discussions based on previous empirical efforts. Specifically, we delineate our analyses from three dimensions: (i) Surrounding existing projects, we study the compatible tools used for DAOs (e.g., on-\&off-chain voting, compatible coordination tools) that can maximally extend the scope of applicability and usage; (ii) Diving into each DAO constructions, we point out several unavoidable drawbacks (e.g., centralization, high cost) that may hinder the DAO progress and development. Chasing an optimal balance-off among all distributed projects should be aligned with concrete requirements. (iii) Excavating historical failures and the reality of today's DAOs (e.g., contract reliance). We further provide several promising directions that can be improved in the future to fit our identified four aspects in results (e.g., multi-DAO Collaboration, usage of subDAOs). 

\begin{figure}[t]
    \centering
   %\ \includegraphics[width=0.99\linewidth]{img/overview.eps}
    \includegraphics[width=0.99\linewidth]{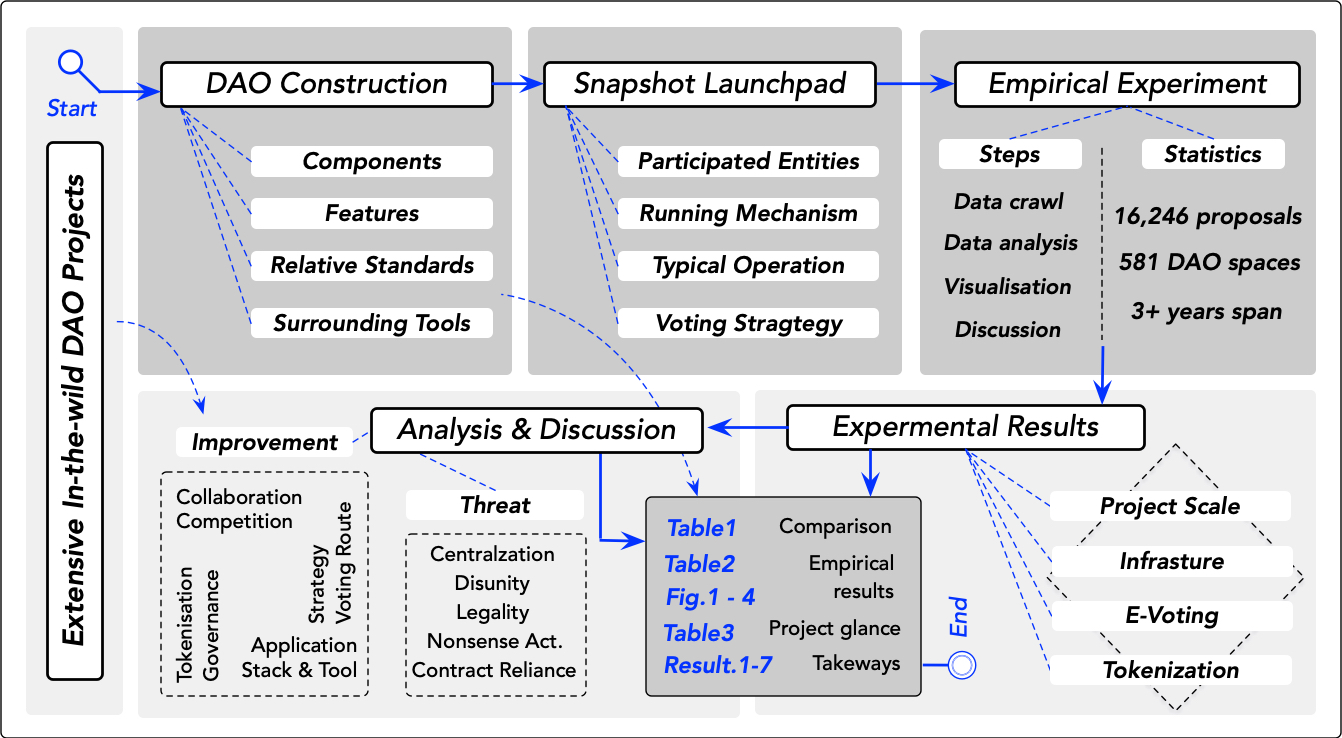}
    \vspace{-0.12in}   
    \caption{Overview of this work}
    \label{fig:9}
    \vspace{-0.23in}
\end{figure}

\smallskip
\noindent\textbf{Key Takeaways.} The rapid growth and widespread adoption of DAOs have brought about significant changes in the way organizations are structured and governed. On the positive side, we observe that DAO participation \&  usage are distributed, application \&  proposal topics are diversified, and execution \& decision-making are automated, which are confirmed by our empirical observations. However, on the flip side, DAO development faces inevitable challenges, including issues of centralization, high costs, unsustainable tokenization mechanisms, and immature supporting technologies (refer to \textcolor{magenta}{Sec.\ref{sec-discuss}} for more information).  All such issues are vital and require much notice. To create a better DAO, we need to examine these issues at every level of the DAO and strive for a healthy approach to distributed governance. This involves developing new tokenization mechanisms that incentivize long-term participation, finding a more fair governance structure, and exploring alternative blockchain technologies that address security concerns.

%The emergence and rapid growth of DAOs have revolutionized organizational structure and governance. On the positive side, the widespread adoption of DAOs has led to a more diverse and inclusive participation, as well as a wider range of applications and proposals being discussed. Additionally, the automation of decision-making and execution has proven efficient and effective, as evidenced by empirical observations. However, despite these benefits, DAOs also face several challenges that must be addressed to improve their efficacy.

%One such challenge is the issue of centralization, which can undermine the decentralized nature of DAOs. High costs associated with DAO development can also limit accessibility and participation, while unsustainable tokenization mechanisms can negatively impact long-term involvement. Furthermore, immature supporting technologies can expose DAOs to security risks and other issues.

%To create a more effective DAO, it is crucial to address these challenges at every level. This includes developing new tokenization mechanisms that incentivize long-term participation, implementing fair governance structures, and exploring alternative blockchain technologies that can enhance security and reliability. 

%Improving the blockchain and smart contracts security is bearing the brunt of reviving the wave of DAO since

%-----------------------------------------
\section{Approaching DAO}
\label{sec-dao}
%-----------------------------------------
In this section, we present a systematic overview of Ethereum DAOs. We achieve this by breaking down the integrated components, identifying key features, reviewing the leading DAOs, and discussing the underlying EIP standards and surrounding tools.

\subsection{DAO Components} 
\label{sec:dao_components}

DAOs consist of various components to facilitate decentralized governance, decision-making, and management. %We list the major components as follows. 

\smallskip
\noindent\textbf{Smart Contract.}
A smart contract is a piece of code that securely runs on blockchain nodes at the same time in a decentralized manner. Thinking of it as a \textit{black-box}, both the input and output are guaranteed synchronized upon reaching a consensus without any assistance of trustworthy third parties. Smart contracts are considered suitable to achieve autonomous organizing by enabling completed self-execution once the defined condition is triggered by traceable and immutable transactions. This enables real-time auditing and verification (e.g., \cite{li2021offline}), hence significantly enhancing the machine-execution security~\cite{tolmach2021survey}. In the context of DAOs, smart contracts are often deployed to create \textit{multi-sig} wallets for secure asset reservation and set \textit{voting strategies} for fair governance.

\smallskip
\noindent\textbf{On-chain Identifier.}
Traditional identifiers that rely on third parties are replaced with Decentralized identifiers (DIDs) \cite{w3cdid} which are not issued, managed, or controlled by any central entity. DIDs are instead managed by individuals whose preferences for data storage platforms are blockchains upon a peer-to-peer  (P2P) network. By making use of Public Key Infrastructure (PKI) technology to generate an asymmetric key pair that is stored on the blockchain, DIDs can achieve globally unique, secure, and cryptographically verifiable authentication services. Typical implementations of DIDs include Ethereum address and Ethereum name service (ENS) \cite{ens}.

\smallskip
\noindent\textbf{Off-chain Snapshot.} 
Snapshot is a technique to record the in-time status of data at a specific height of blocks. Such a type of technique is quite important for DAO governance where all the historical results voted by participants are recorded as evidence, which is necessary for both on- and off-chain governance. Off-chain signatures are often used to adjust the weights of on-chain tokens during the voting process. To achieve a smooth collaboration, a snapshot of on-chain balances and addresses will be captured to determine voting rights, and then the participants of community members will start to vote for DAO proposals under the weights. In this way,  on-chain transaction fees are significantly waived. Notably, the name of the Snapshot platform, researched in this paper, exactly comes from this technical term.

\smallskip
\noindent\textbf{Stake/Governance Token.}
The self-controlled, and portable DIDs can offer tamper-proof and cryptographically secure attestations for on-chain decentralized identity. By raising the burden on each attestation's provenance and validity to be securely proved, at the same time, easing the validation process, DIDs become suitable for implementing wallet services in which stakes and utility tokens can be securely stored. Stakes refer to the tokens that a holder can deposit in the system. The more stakes a holder provides, the higher confidence he will have in operating consensus procedures (e.g., Proof-of-Stake). In contrast,  utility tokens are designed to be used for a specific purpose, especially in a DApp or in a game. They offer users benefits such as access to products and services. Staked tokens and utility tokens, in most cases, are separate where the former ensures the normal operation of systems, and the latter is used for governance \cite{kiayias2022sok} and votes in the context of DAO. In this sense, communities sometimes equivalently use the name of \textit{governance token}. Besides, the staked tokens can be further used to establish an on-chain reputation, which is primarily to give corresponding values to individuals who frequently participate DAOs.

\smallskip
\noindent\textbf{Reputation Mechanism.} Reputation is a crucial element in maintaining trust and promoting collaboration within DAOs. It serves as a measure of a member's contributions, determining their level of influence. Members can earn a reputation by actively participating in governance decisions, providing liquidity to a protocol, or contributing to ongoing projects. The more a member contributes to a DAO, the higher their reputation will be. This reputation can be leveraged in various ways, such as determining voting power in governance decisions, allocating rewards from the organization's treasury, or granting access to certain resources and privileges. Typically, reputation is quantified through the number of governance tokens held by a member.

\smallskip
\noindent\textbf{Secure e-Voting Scheme.}
Although traditional e-Voting systems have been growing, they are still susceptible to manipulation. One of the most critical problems is it being prone to the Sybil attack~\cite{sybil} where malicious users create false identities to vote. In the DAO space, by using DIDs and asking for an on-chain attestation, the integrity of the e-Voting process could be improved. Staked tokens and utility tokens, which are bound with DIDs, are also commonly used in e-Voting to represent the voting influence. Based on our investigation, existing DAO voting schemes are based on relatively simple mechanisms such as \textit{basic voting}, \textit{single-choice voting} and \textit{ranked choice voting} (cf. \textcolor{teal}{Fig.\ref{fig:4}}), rather than complicated cryptographic e-Voting systems \cite{cortier2016sok}.

\subsection{DAO Features}\label{subsec-framework}

We examine DAOs from four key perspectives: operational mechanism (for underlying foundations/dependencies), functional features (processing phases), non-functional features (advanced properties), and market performance (real-world impact). We additionally investigate 30+ projects summarised in \underline{Table~\ref{tab-summary}}.

\smallskip
\noindent\textbf{Operational Mechanism.}
\ding{172} \textit{Network} refers to the underlying blockchain platform on which the DAO operates. Since DAOs rely on self-executing contracts where the terms of the agreement are directly written into the code, the network plays a crucial role in determining the functionality and shape of these smart contracts. \ding{173} \textit{Protocol/Field} describes the specific usage or application of the organization. This can range from areas such as finance and governance to art and social impact. \ding{174} \textit{Governance token} represents the voting power in DAO governance. Stakeholders can thereby vote on proposals to make decisions and allocate resources. 

\smallskip
\noindent\textbf{Functional Features.} Based on the DAO projects on Snapshot, we conclude that a typical lifecycle of DAOs includes the phases of \textit{create}, \textit{propose}, \textit{vote}, and \textit{action}. Specifically, \ding{172} \textit{create} involves setting up the initial configurations of the DAO, covering not only the DAO space and related information but also personal identifiers (e.g., ENS, DiD). \ding{173} \textit{Propose} focuses on drafting, editing, and releasing proposals. Specific requirements will be applied to the proposer, such as holding enough stakes. \textit{vote} \textit{Vote} calls for feedback and preferences from community participants. Stakeholders can vote for multiple options, mostly just \textit{for} or \textit{against} based on their interest and willingness. Different voting strategies will be used to adjust the power of a voter. Further, \textit{action} is to execute the decisions once reaching an agreement. Although this phase is critical, it cannot be effectively measured, so we have omitted it from our table.

\smallskip
\noindent\textbf{Non-functional Features.} \ding{172} \textit{Permissionless} is a key factor to measure decentralized governance due to its dynamic joining/leaving mechanisms. Token holders can make decentralized decision-making by voting on preferred proposals and influencing the organization's direction. \ding{173} \textit{Transparency/Immutability} means that all transactions and decisions within a DAO will be transparent and cannot be tampered with, fostering trust among members and stakeholders. \ding{174} \textit{Anti-censorship} refers to the ability to prevent stakeholders from censoring the flow of transactions (e.g., OFAC compliant).  \ding{175} \textit{Interoperability} is the feature that allows a DAO to interact and exchange data with other DAOs, enabling seamless integration and collaboration with various ecosystems. \ding{176} \textit{Token-based Incentives.} Token-based incentives align the interests of stakeholders and encourage active participation. Members can earn tokens by contributing to the organization or by staking them to support projects, resulting in a more engaged community.

\smallskip
\noindent\textbf{Market Performance.} Market performance can be evaluated via quantitative metrics in multiple dimensions. \ding{172} \textit{Treasury} refers to a pool of funds that is collectively owned and controlled by the members of the organization. With similarity to the total value locked (TVL) in DeFi, the treasury is typically built up over time through contributions from members or profits generated from the organization's activities. \ding{173} \textit{Holders} represents participants who own governance tokens and are therefore eligible to vote on proposals that shape the direction of the organization. It can provide insight into the level of participation and engagement within the DAO. \ding{174} \textit{Proposals} are the specific documents that outline a suggested course of action for the organization. These proposals can be put forward by any member of the DAO. \ding{175} \textit{Votes} counts for the total number of votes (equiv. decisions) cast by stakeholders.

\smallskip
\noindent\textbf{A Collection of Leading DAOs} (\textit{upper} \underline{Table~\ref{tab-summary}}). Based on the aforementioned metrics, we investigate a group of (30+) DAO projects that are currently active and operating in real-world environments. These selected projects are highly influential within their respective communities, as evidenced by their market performance. It's worth noting that most of the selected DAOs operate on the Ethereum blockchain (also supported by \textcolor{teal}{Fig.\ref{fig:3}}) and are classified as belonging to the DeFi track (by \textcolor{teal}{Fig.\ref{fig:1}}). Additionally, DAOs are expected to have certain essential properties such as permissionless access and transparency, but the others possess additional qualities.

\subsection{Supporting Standards}

Recall that the standards referred to in this paper are formatted in technical documents dedicated to on-chain programming. Conventions are established by using the standards during programming without having to reinvent the wheel, making it easier and more efficient for applications and contracts to interact with each other. Here, we list the relevant standards that support DAO scenarios. 

\smallskip
\noindent\textbf{\text{EIP-20/BEP-20.}} Common Interfaces for fungible-token (FT)~\cite{eip20}. Running the e-voting normally requires stakes and utility tokens that typically implement ERC-20 on Ethereum, BEP-20 on BSC, or similar standards on any other blockchain platforms.

\smallskip
\noindent\textbf{\text{EIP-721/BEP-721/EIP-1155.}}  Common Interfaces for the non-FT (NFT) \cite{eip721} and multi-token \cite{eip1155}. Stakes and governance tokens can also include the forms of NFTs \cite{wang2021non}, being the voting power of e-voting or being the deposit for users to participate in any campaigns of a DAO. BEP means the standards of BSC.

\smallskip
\noindent\textbf{\text{EIP-4824.}} Common Interfaces for DAOs \cite{eip4824}. This standard aims to establish conventions on matching on- and off-chain representations of membership and proposals for DAOs by using \textit{daoURI}, an indicative reference inspired by ERC-721, which enhances DAO search, discoverability, legibility, and proposal simulation.

\smallskip
\noindent\textbf{\text{EIP-1202.}} Common Interfaces for the voting process \cite{eip1202}. The standard implements a range of voting functions (e.g., $\mathsf{VoteCast}$, $\mathsf{castVote}$, $\mathsf{MultiBote}$) and  informative functions (e.g., voting period, eligibility criteria, weight) to enable on-chain voting as well as to view the vote result and set voting status.

\smallskip
\noindent\textbf{\text{ERC-779.}} Common Interfaces for DAOs \cite{eip779}. Unlike other hard forks that have altered the Ethereum protocol, the DAO Fork is executed solely through the alteration of the state of the DAO smart contract whereas transaction format, block structure, and protocol were not changed. It is an "irregular state change" that was transferred ether balances from the child DAO contracts into a specified account, the "WithdrawDAO" contract.

\subsection{Surrounding Tools}

Additionally, a wide variety of tools have been proposed to ease the process of joining, launching, and managing a DAO. We list several of them in the \textit{bottom} of \underline{Table~\ref{tab-summary}}. Besides the launchpads that can manage DAOs, a host of providers introduce their services and infrastructure \cite{daoreport} such as token services (e.g., MakerDAO for maintaining the DAI stablecoin),  on\&off-chain voting tools (Tally, Snapshot), treasury oversight (TokenTerminal, Zapper), growth products, risk management (Gnosis), task collaboration (Mirro, Colony), community platforms (MolochDAO, Metagovernance), analytic tools (Dune, RootData), operational tools (Aragon, DAOstack), wallet services (Gnosis Safe) and legal services (LegalDAO).

%-----------------------------------------
\section{Diving into Snapshot}
\label{sec-snapshot}
%-----------------------------------------

% \subsection{Overview of Snapshot}
Snapshot is an off-chain voting system designed for DAOs created on multiple blockchain platforms. The system has been widely adopted by many crypto startups and companies  to assist in surveying users. Each project can create proposals for users to poll votes by using the staked or governance tokens. All the voting procedures are essentially feeless as the operation is executed off-chain, avoiding costly on-chain verification. Users only need to connect their wallet to the launchpad and allow the action of signing. Besides, the projects, voting proposals for each project, and corresponding results are stored based on the IPFS decentralized storage system \cite{benet2014ipfs}. The snapshot thereby becomes a convenient tool for DAO creators to query the feedback from communities and audiences. Here, we provide detailed actions for each party. 

\begin{itemize}
    \item \textit{DAO creator.} DAO creators are those companies or projects that aim to use Snapshot. The creator needs to hold a valid ENS domain and register his project on the Snapshot launchpad by creating a profile with inputs of detailed information such as project \textit{name}, \textit{about}, \textit{website}, \textit{symbol}, \textit{service}, \textit{network} (equiv. blockchain platforms) and contacts like \textit{Twitter}, \textit{Github} and \textit{CoinGecko}. 

    \item \textit{Poll proposer.} They can create their proposals for a specific project if he holds a sufficient amount of relevant governance tokens. In many cases, poll proposers are the DAO-creating team members as they have enough staked tokens and motivations to improve the protocol.
    
    \item \textit{Users.} Users can vote for each proposal based on their preferences. All participants are required to have valid accounts with staked tokens, such as an Ethereum address or a short name registered on ENS. Users can add a record on accounts to allow votes to be viewable at the connected addresses. 
\end{itemize}

%\smallskip
\noindent\textbf{Running Mechanism.} Snapshot roots in the technique of \textit{snapshot}. The snapshot technique is to record the in-time token-holding status of all accounts and wallets on-chain at a specific block height. It acts as the way of a camera, taking photos of the entire picture at the moment. In this way, a stakeholder can learn information like \textit{who has the token}, \textit{how many tokens they have}, etc. Owing to the benefits of transparency and traceability, the technique has been applied to many crypto-events, such as \textit{airdrops} for incentive distribution and \textit{compensation} for users after hacking or attacks. Accordingly, the Snapshot project leverages such technology to solve the problem existing in the voting processes. It can intercept the historical data at a certain block height and the associated holding status (e.g., accounts, tokens, NFTs) of a certain type of token. Based on these data, the voting weights can be reasonably assigned to individual community members aligned with different rules.

\smallskip
\noindent\textbf{Typical Operations.} Based on different roles, we capture three main types of operations. Notably, operations on Snapshot are aligned with the DAO deployed on other launchpads.

\begin{itemize}

\item \textit{Creating spaces.} If a project aims to introduce decentralized governance into the project, they can create a Space in Snapshot for users to propose proposals and perform voting processes. As discussed, a distributed identifier is required before the application. This identifier is used to connect the created unique project profile. The community (equiv. $\mathsf{Space}$) is created once the basic information is fully fulfilled. Importantly, setting the community's distribution strategy (a.k.a., $\mathsf{Strategy}$). It is written a Javascript function that can be used to adjust the weight of impact.

\item \textit{Proposing proposals.} To propose a proposal in the community, the community member must first adhere to the guidelines set forth by the community manager. For instance, an eligible user in the ENS community is required to hold more than 10k ENS for creating proposals. Once meeting these requirements, the proposer can fill in the proposal content, options, start/end dates, and set voting rules.

\item \textit{Poll/Vote.} The voting process is open for the community only if a user has governance tokens. Every project has its unique governance tokens where a user can even trade (buy/sell/exchange) them on secondary markets. The voting process is designed in a clean and simple style: connect to the wallet, select options, and sign with signatures. Users can view their options, voting power, and snapshot time for each submission of voting. All the data is obtained from the snapshot. 

\end{itemize}

%\smallskip
\noindent\textbf{Voting Strategy.} As the most essential part of profit distribution, different strategies provide a series of methods of calculating voting power. The strategy in Snapshot is essentially a JavaScript function. Users can combine at most 8 strategies on every single proposal while voting power is cumulative. Meanwhile, users can write customized strategies according to their requirements. At the time of writing, Snapshot has 350+ voting strategies and \textit{ERC20-balance-of} is the most adopted strategy. We list mainstream strategies here.

\begin{itemize}

\item \textit{Delegated voting.} The voting power is based on a delegation strategy. Only permitted stakeholders have valid impacts on the voting process. 

\item \textit{Weighted voting.} The voting power can be calculated either by the \textit{single} weight strategy (one-coin-one-vote) or a \textit{quadratic} strategy. The quadratic strategy weakens the significant influence of rich stakeholders, diminishing the gap across different individuals.

\item \textit{Whitelist voting.} The permitted stakeholders who are on the whitelist are allowed to vote. The whitelist may either get updated manually or by certain rules. 

\item \textit{NFT voting.} Voting by using NFT needs to be compatible with ERC-721 or ERC-1155 based strategies.

\end{itemize}

%\footnote{Source: \url{https://snapshot.org/\#/?type=strategies}. [Nov 2022]}

%-----------------------------------------
\section{Experiments and Results}
\label{sec-experi}
%-----------------------------------------

This section provides our experiments and corresponding results. %We detail our methods as follows. 

\vspace{-0.1in}
\begin{table}[!hbt]
\centering
\caption{\textcolor{teal}{Result} Guidance}\label{tab-result}
\vspace{-0.15in}

\resizebox{\linewidth}{!}{
\begin{tabular}{llr}
\toprule
& \multicolumn{1}{c}{\textit{\textbf{Index}}} & \multicolumn{1}{c}{\textit{\textbf{Description}}}  \\ 
\midrule
\multirow{5}{*}{\rotatebox{90}{\textit{\textbf{Project scale}}}} &   \textcolor{teal}{Fig.\ref{fig:1}} & Number of registered members in each considered DAO project.  \\
& \textcolor{teal}{Fig.\ref{fig:10}} & Number of votes of each proposal among all considered DAO projects. \\
& \textcolor{teal}{Fig.\ref{fig:5}} &  Up-to-date number of DAO projects kicked off every month. \\
& \textcolor{teal}{Fig.\ref{fig:6}} &  Duration of each proposal among all considered DAO projects.   \\
& \textcolor{teal}{Fig.\ref{fig:12}} & Languages distribution among all considered DAO projects.   \\ 

\midrule

\multirow{3}{*}{ \rotatebox{90}{\textit{\textbf{Infrast.}}}} & \multirow{1}{*}{\textcolor{teal}{Fig.\ref{fig:3}}} & \makecell[r]{Fraction of different blockchain networks being used for \\ running each considered DAO projects.}  \\
& \multirow{1}{*}{ \textcolor{teal}{Fig.\ref{fig:11}}}  &  \makecell[r]{Fraction of different IPFS addresses being used for data storage \\ of each proposal among all considered DAO projects.} \\

\midrule

\multirow{5}{*}{ \rotatebox{90}{\textit{\textbf{e-Voting}}}} & \multirow{1}{*}{\textcolor{teal}{Fig.\ref{fig:4}}} & \makecell[r]{Fraction of different voting mechanisms being used for e-voting of \\  each proposal among all considered DAO projects.}  \\
& \multirow{1}{*}{\textcolor{teal}{Fig.\ref{fig:7}}}  &  \makecell[r]{Voting patterns (in terms of the number of candidates and variances} \\
& \multirow{1}{*}{\,\, \&\textcolor{teal}{\ref{fig:8}}}  & of results) among all considered DAO projects. \\
& \multirow{1}{*}{\textcolor{teal}{Fig.\ref{fig:9}}}  &  \makecell[r]{Number of votes of each proposal among all considered DAO projects.} \\
& \multirow{1}{*}{\textcolor{teal}{Fig.\ref{fig:14}}}  &  \makecell[r]{Clustering among all considered DAO projects.} \\

\midrule

\multirow{3}{*}{ \rotatebox{90}{\textit{\textbf{\makecell{Token \\Usage}}}}} & \multirow{1}{*}{\textcolor{teal}{Fig.\ref{fig:2a}}} & \makecell[r]{Fraction of the usage of prevalent DAO tokens and other \\  self-issued tokens in the Snapshot.}  \\
& \multirow{1}{*}{\textcolor{teal}{Fig.\ref{fig:2b}}}  &  \makecell[r]{Fraction of the usage between different prevalent DAO tokens.} \\
& \multirow{1}{*}{\textcolor{teal}{Fig.\ref{fig:2c}}}  &  \makecell[r]{Fraction of the usage between different self-issued DAO tokens.} \\

\bottomrule
\end{tabular}
}
\vspace{-0.15in}
\end{table}

\smallskip
\noindent\textbf{Measurement Establishment.} Our experiment consists of three steps. Firstly, we develop a crawling script, which is deployed on AWS EC2 cloud server ($\mathsf{m6i.32xlarge}$) with 128-vCPU and 512GiB-memory, to capture all data from the Snapshot platform. The script is designed to collect all the information presented on the Snapshot main page and subpages created by DAO creators, including numerical values (such as voting results and participation scales) and context-aware strings (such as language usage and topic classification). All data will be compiled into a final CSV document. Secondly, we analyze the data using Python and generate corresponding visualizations. During this analysis, we sort, clean up, and classify the metadata to obtain meaningful results. Finally, we present our findings in this section and provide derived insights.

\smallskip
\noindent\textbf{Overall Statistics.} Our study is based on the analysis of 16,246 proposals from the 581 most prevalent DAO projects collected from Snapshot. Our data crawling covers a period of more than three years, starting from the establishment of Snapshot in August 2020 and continuing until November 2022 (the time of writing this paper). We have collected a wide range of key data fields such as project title, number of members, proposal title, status, network, IPFS address, voting strategy, project start/end date, block snapshots, result name, result outcome, proposal content, number of votes, etc. To present our statistical results clearly, we have categorized them into four perspectives, namely, \textit{project scale}, \textit{infrastructure}, \textit{e-voting schemes}, and \textit{token usage}. To guide the readers through the result analyses, we provide a high-level overview in \underline{Table~\ref{tab-result}}.

\smallskip
\noindent\textbf{Data Use Disclaimer.} All the data we crawl from Snapshot are open-released and free to use with CC0 licenses. We strive to maintain the accuracy of all data that we crawl from Snapshot and declare that the data will not be used for any commercial purposes.

%%%%-----------------------------------
\subsection{Project Scale}

\begin{figure}[t]
\centering
\subfigure[Number of members in different DAOs]{
\resizebox{0.48\textwidth}{!}{
    \includegraphics[width=0.85\linewidth]{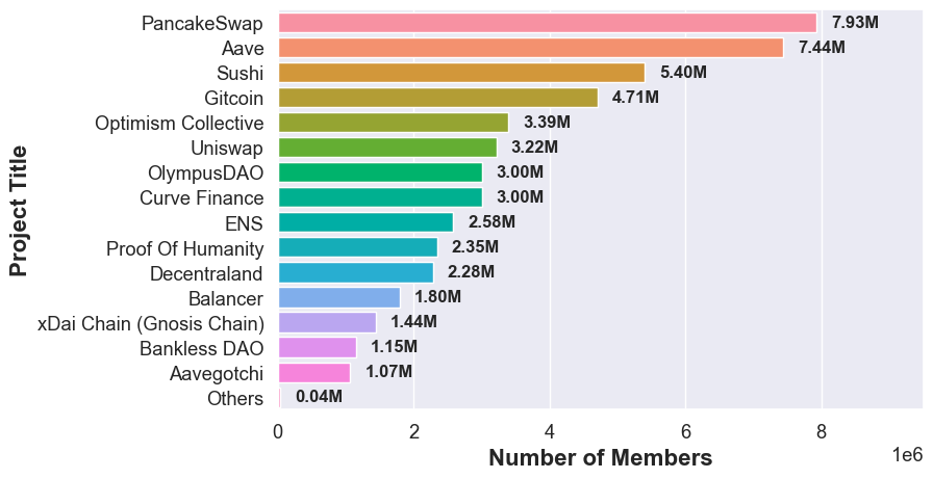}
\label{fig:1}
}
}
% \subfigure[Voting patterns with the number of candidates]{
% \resizebox{0.37\textwidth}{!}{
%   \rotatebox{0}{ \includegraphics[width=1\linewidth]{img/fig-9.pdf} }
% \label{fig:9}
% }
% }

%\hspace{2em}
\vspace{-0.1in}
\subfigure[DAO launching dates]{
\resizebox{0.48\textwidth}{!}{
    \includegraphics[width=0.85\linewidth]{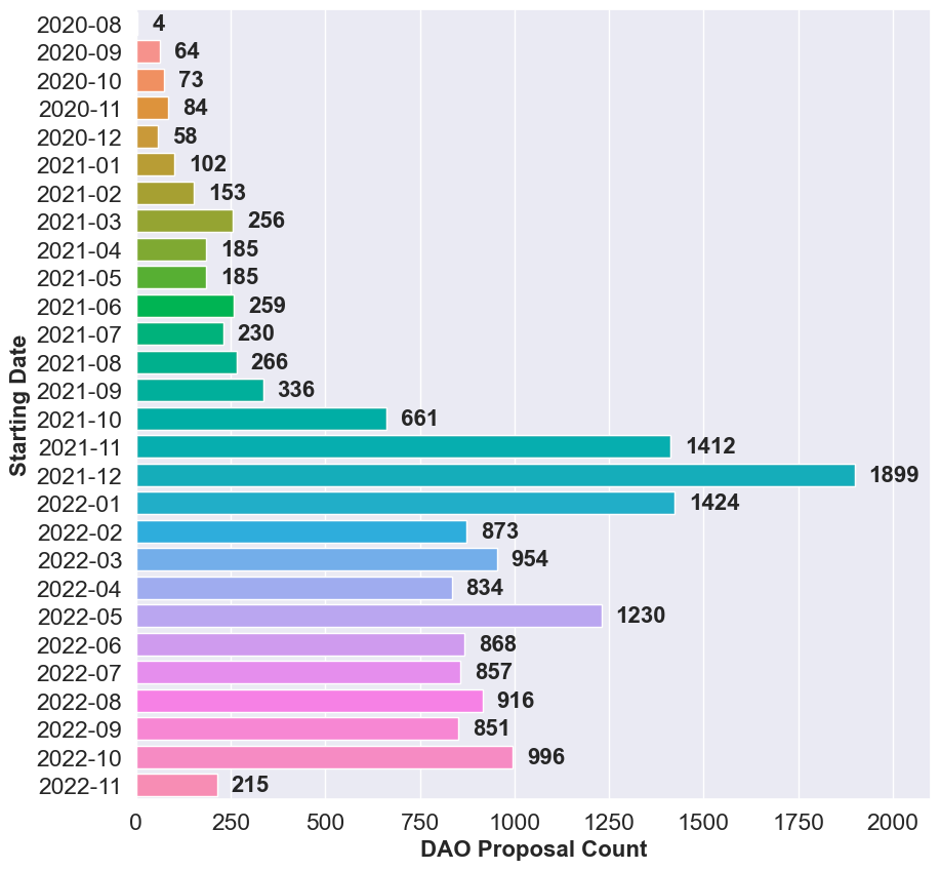}
\label{fig:5}
}
}
\vspace{-0.15in}
\caption{Project Scale}
\label{fig:1:bar}
\vspace{-0.2in}
\end{figure}

This sector describes the scale of the considered DAO projects in views of its \textit{participating members}, \textit{vote distributions} for proposals, \textit{launching dates}, \textit{active proposal duration}, and \textit{language distribution}. %We provide details of each item.

\textcolor{teal}{Fig.\ref{fig:1}} shows that the very top DAO projects have reached a six-order of magnitude (millions), with the two most popular projects reaching over 7M members, i.e., \textsc{PancakeSwap} and \textsc{Aave}. Note that the ``Others'' bar computes the average number of members of all the DAO projects from the 16th and onwards. This indicates that the member distribution is likely following the Pareto principle \cite{pareto} that the majority of DAO participants show up to a small part of DAO communities. When diving into each DAO proposal, it can be found from \textcolor{teal}{Fig.\ref{fig:10}} that the fraction of having over 100 votes and having less than 10 votes come first and second, respectively. This also indicates the Pareto principle is being complied with in the sense that a huge number of votes are aggregated to a small portion of proposals, while there are still a significant number of proposals that are marginalized by the community.

\textcolor{teal}{Fig.\ref{fig:5}} shows that the concept of DAO appears to be accepted and realized by a broader public since Q3 2020 (align with \cite{daoreport}). From then on, the Web3 supporters kept drawing traffic to the DAO community. It turns out that a peak arose from Nov 2021 to Jan 2022 in regard to the number of projects being kicked off during the period (along with the booming of DeFi and NFT). It can also be found that the average monthly number of new projects is much higher than that before the peak, which indicates a milestone in the development of DAO communities. According to \textcolor{teal}{Fig.\ref{fig:6}}, the duration of each proposal is found mostly within a week, which is reasonable and is matched with the duration of many real-world election campaigns. \textcolor{teal}{Fig.\ref{fig:12}} shows that English is the most popular language used in the proposals, accounting for 75.1\% of the proposals among all considered DAO projects in our collection. Chinese comes second with a fraction of 4.3\%, followed by Germany (2.9\%), Korean (1.8\%), Italian (1.5\%), and French (1.4\%). All the rest of the languages are categorized in ``Others'', accounting for 12.9\% of the proposals. The usage of languages can also indirectly reflects the nationality distribution of participating members.

\begin{center}
\tcbset{
        enhanced,
      % colback=black!5!white,
        boxrule=0pt,
      %  colframe=black!75!black,
        fonttitle=\bfseries
       }
       \begin{tcolorbox}[
     %  title=Taking-home Messages,
       lifted shadow={1mm}{-2mm}{3mm}{0.1mm}{black!50!white}
       ]
      \textbf{\text{Insight-\ding{202}:}} The DAO community has been tremendously developed across many countries and regions and the usage has remained high. However, the member distribution follows the Pareto principle (the 80-20 rule \cite{pareto1964cours}), which requires further notice to avoid unexpected centralization.
       
       \end{tcolorbox}
\end{center}

%%%%-----------------------------------
\subsection{Infrastructure}
This sector describes the information of \textit{IPFS network} and \textit{blockchain platform} infrastructure being used by the considered DAO projects.% or proposals, followed by details here. 

\begin{figure*}[htb!]
\centering
\subfigure[Number of votes of each proposal]{
\resizebox{0.23\textwidth}{!}{
    \includegraphics[width=0.95\linewidth]{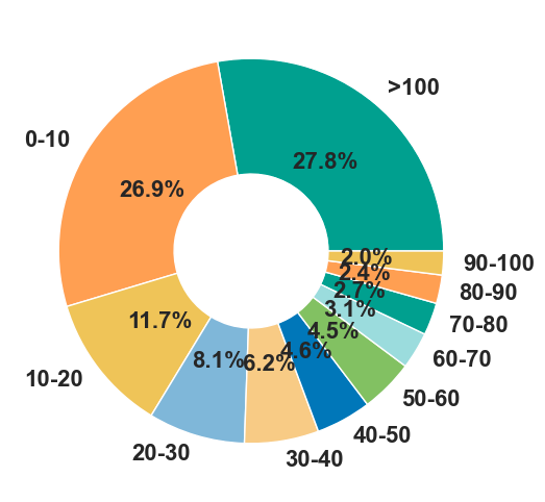}
\label{fig:10}
}
}
%\hspace{2em}
\subfigure[Duration of each proposal]{
\resizebox{0.22\textwidth}{!}{
    \includegraphics[width=0.95\linewidth]{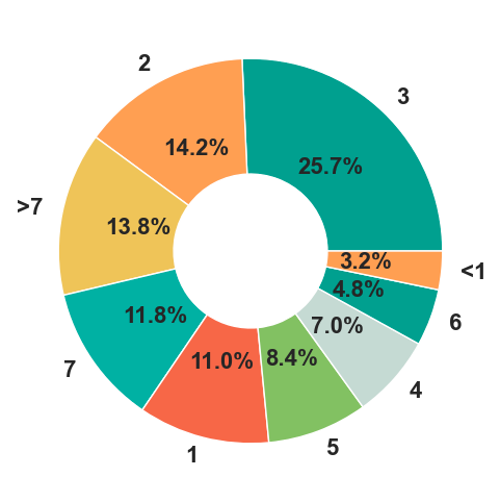}
\label{fig:6}
}
}
%\hspace{2em}
\subfigure[Proposal voting variances]{
\resizebox{0.23\textwidth}{!}{
    \includegraphics[width=0.95\linewidth]{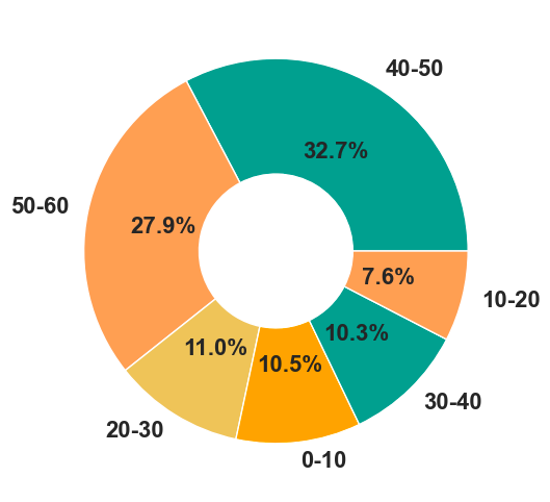}
\label{fig:7}
}
}
%\hspace{2em}
\subfigure[Project voting variances]{
\resizebox{0.22\textwidth}{!}{
    \includegraphics[width=0.95\linewidth]{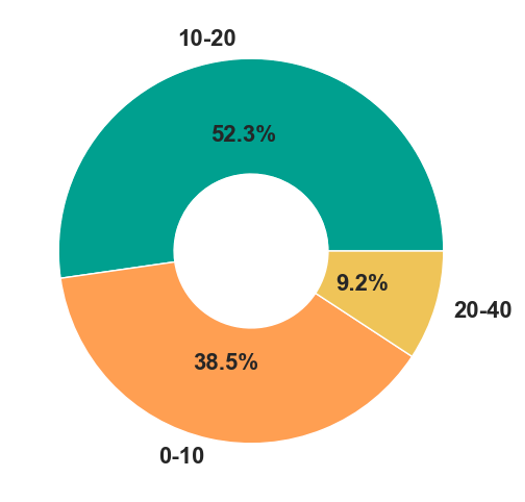}
\label{fig:8}
}
}
\subfigure[Language distribution]{
\resizebox{0.18\textwidth}{!}{
    \includegraphics[width=0.95\linewidth]{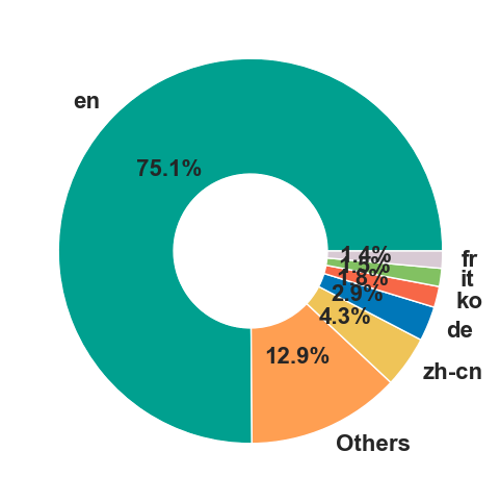}
\label{fig:12}
}
}
\subfigure[Fraction of e-voting schemes]{
\resizebox{0.31\textwidth}{!}{
    \includegraphics[width=0.95\linewidth]{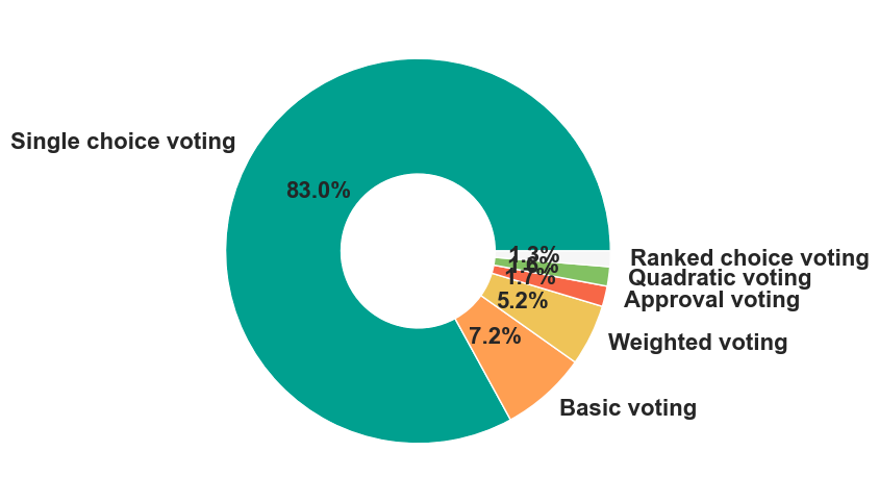}
\label{fig:4}
}
}
\subfigure[Fraction of blockchains]{
\resizebox{0.26\textwidth}{!}{
    \includegraphics[width=0.95\linewidth]{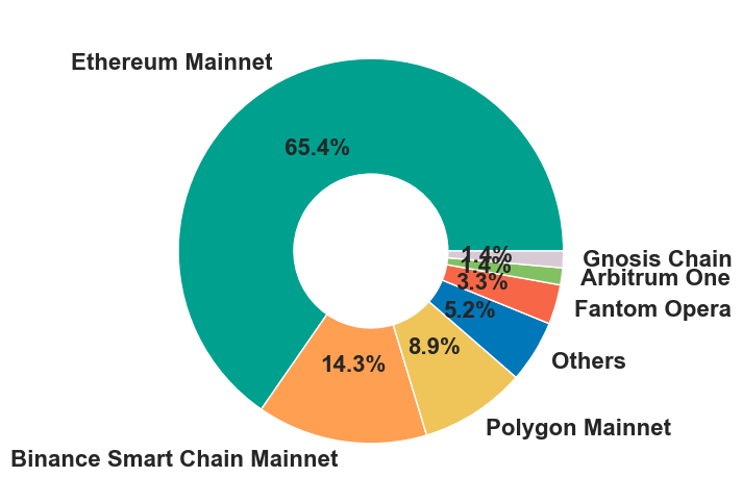}
\label{fig:3}
}
}
%\hspace{2em}
\subfigure[Fraction of IPFS storage]{
\resizebox{0.175\textwidth}{!}{
    \includegraphics[width=0.93\linewidth]{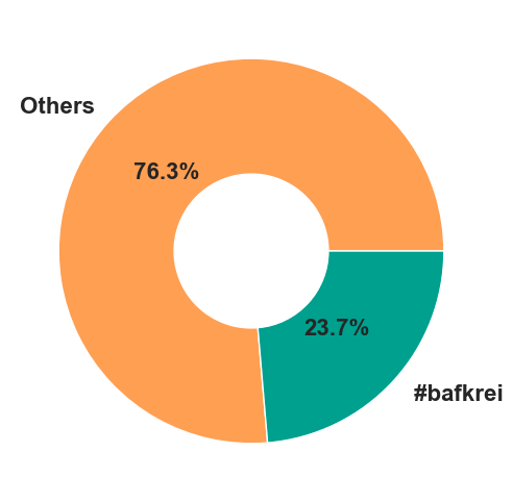}
\label{fig:11}
}
}
\caption{Snapshot Empirical Results in Multi-dimensions}
\label{fig:2:pie}
\end{figure*}

It is found from \textcolor{teal}{Fig.\ref{fig:3}} that Ethereum Mainnet~\cite{wood2014ethereum} is the most popular blockchain platform used by the considered DAO projects in our collection, accounting for a fraction of 65.4\%. Binance Smart Chain Mainnet comes second with a fraction of 14.3\%, followed by Polygon Mainnet (8.9\%), Fantom Opera (3.3\%), Arbitrum One (1.4\%), and Gnosis Chain (1.4\%). All the other platforms are categorized in ``Others'', accounting for 5.2\% of the projects. In regard to the decentralized IPFS data storage as illustrated in \textcolor{teal}{Fig.\ref{fig:11}}, the ``\#bafkrei'' is used the most often, accounting for 23.7\% among all the proposals. This implies that 76.3\% of the DAO proposals (starting with ``\#QM'') are still using tools like nmkr.io or other minting platforms that use the outdated version of content identifier (CIDv0~\cite{ipfscid}, Base58) which is more expensive and less effective for IPFS data storage. This deserves a caution that there are lack of motivation for DAO developers to upgrade their infrastructure, which might degrade the capacity and efficiency of data storage to DAO communities if IPFS would only completely root for CIDv1~\cite{ipfscid} in the future.

% Here, we try to explore such an interesting address to see why it occupies a significant portion of the whole page.
% Nevertheless, based on our investigation, the excessive reliance on single-point data storage leads to centralization, posing a threat of the single point of failure to the DAO community. This issue, in our view, requires further improvements.

\begin{figure}[t]
    \centering
    \includegraphics[width=0.9\linewidth]{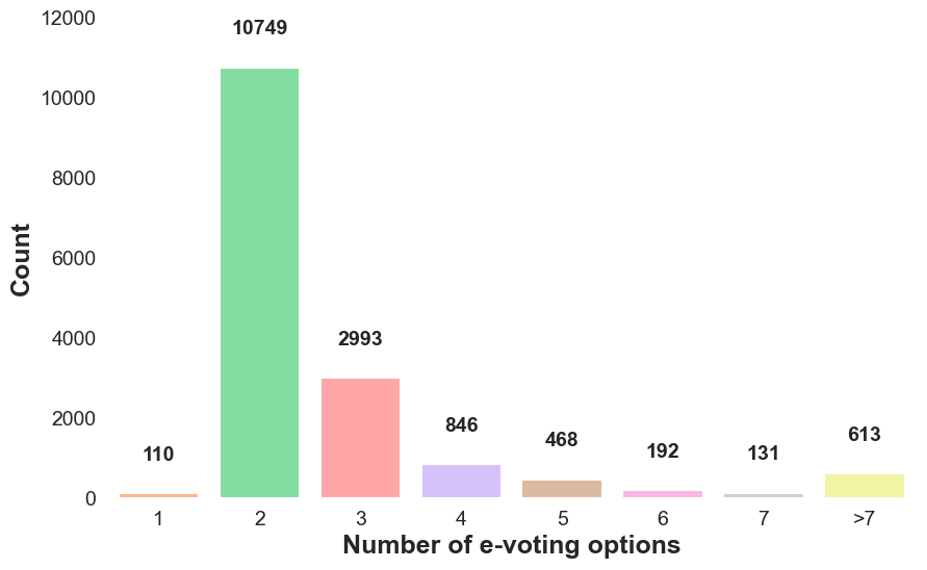}
    \caption{Voting patterns with the number of candidates}
    \label{fig:9}
\end{figure}

\begin{center}
\tcbset{
        enhanced,
      % colback=black!5!white,
        boxrule=0pt,
      %  colframe=black!75!black,
        fonttitle=\bfseries
       }
       \begin{tcolorbox}[
     %  title=Taking-home Messages,
       lifted shadow={1mm}{-2mm}{3mm}{0.1mm}{black!50!white}
       ]
      \textbf{\text{Insight-\ding{203}:}}  It is fortunate that the platforms used by existing DAO proposals and projects are diversified. On the contrary, the insufficient motivation of upgrading the content identifier version of IPFS data storage may degrade the capacity and efficiency of the community. 
       
       \end{tcolorbox}
\end{center}

%%%%-----------------------------------
\subsection{E-voting Scheme}
This sector describes \textit{the number of valid votes} and \textit{the fraction of different e-voting schemes} used in the considered DAO projects and proposals, as well as \textit{different voting patterns}.

\smallskip
\noindent\textbf{E-voting schemes.} It is found from \textcolor{teal}{Fig.\ref{fig:4}} that the e-voting schemes can be categorized into the following ranks. The single-choice voting is the dominant strategy that accounts for 83.0\% among all reviewed strategies, followed by the basic voting strategy with a fraction of 7.2\%.  Conversely, the weighted voting (5.2\%), approval voting (1.7\%), quadratic voting (1.6\%), and ranked choice voting (1.3\%) strategies are rarely adopted by participating members in comparison. The results demonstrate that single-choice voting is the most popular strategy. This indicates that DAO users still prefer to adopt the simplest way of polling. Although an intuitive concern comes that the single-choice voting and basic voting strategies may result in Matthew effect \cite{matthew} in vote distribution, the results show that most proposal advisors ignore such drawbacks in practice. 

%\begin{itemize}
%\item \textbf{Single-choice voting. }(83.0\%) XXXX
%\item \textbf{Basic voting. }(7.2\%) XXXX
%\item \textbf{Weighted voting. }(5.2\%) XXXX
%%\item \textbf{Approval voting. }(1.7\%) XXXX
%\item \textbf{Quadratic voting. }(1.6\%) XXXX
%\item \textbf{Ranked choice voting. }(1.3\%) XXXX
%\sy{(@Qin: which is better?? Any risks to having the single-choice voting as the majority??)}

\smallskip
\noindent\textbf{Voting Patterns.} \textcolor{teal}{Fig.\ref{fig:9}} shows that the most popular voting pattern in our collection is the binary voting with a quantity of over 10,000 proposals, followed by the ternary and quaternary voting patterns. On the other hand, \textcolor{teal}{Figs.\ref{fig:7}} and \textcolor{teal}{Figs.\ref{fig:8}} learn the variances of voting results of each proposal and each project, respectively, aiming to investigate how much a statement is being agreed with or opposed to by the community. It is realized from \textcolor{teal}{Fig.\ref{fig:7}} that over 60\% of the proposals end up with a large variance of over 40. This indicates that any e-voting held in the current DAO communities is more likely to end up with a one-sided result. In each project, nevertheless, the number of projects with an average great variance is not that large, only accounting for 9.2\%. Balanced results account for 38.5\% while there are 52.3\% of the projects have an average variance ranging from 10 to 20. This indicates that the voting pattern of a one-sided result rarely happens in a project-wised context. Each project may have several one-sided voting results with the majority more likely ending up with a balanced result.

Here, we try to explain the reason that causes the variance differences between proposal- and project-level voting results. As observed in \textcolor{teal}{Fig.\ref{fig:9}}, most of the voting results are binary-based patterns, whose corresponding variances are naturally very large. This will significantly increase the result (value) of proposal-level variances as each proposal is merely established on top of one voting pattern. In contrast, the results in project-level variances are relatively balanced because each project contains a series of proposals that may moderate the extreme value caused by binary results. In our view, one-sided results do not necessarily mean ``bad'', which instead indicate that DAO members are more likely to make an instant decision without significant debates. A balanced result shows that DAO communities are difficult to reach an agreement among the participants. However, on the flip side, this exactly reflects the so-claimed properties of  \textit{decentralization} or \textit{democracy}. Controversial arguments indicate that defining \textit{what is a normal or healthy voting result} is complicated in an unclear context. 

\begin{center}
\tcbset{
        enhanced,
      % colback=black!5!white,
        boxrule=0pt,
      %  colframe=black!75!black,
        fonttitle=\bfseries
       }
       \begin{tcolorbox}[
     %  title=Taking-home Messages,
       lifted shadow={1mm}{-2mm}{3mm}{0.1mm}{black!50!white}
       ]
      \textbf{\text{Insight-\ding{204}:}} Current e-voting patterns and results of many DAO projects reflect the decentralization or democracy in DAO communities, but at the same time bringing difficulty in reaching an agreement, at least not as effective as traditional e-voting does. This is essentially the weakness of the flat organizational structure, where the trade-off between the flat- and hierarchical-structure remains debatable.
       
       \end{tcolorbox}
\end{center}

\vspace{0.5em}

\begin{figure}[t]
    \centering
    \includegraphics[width=0.9 \linewidth]{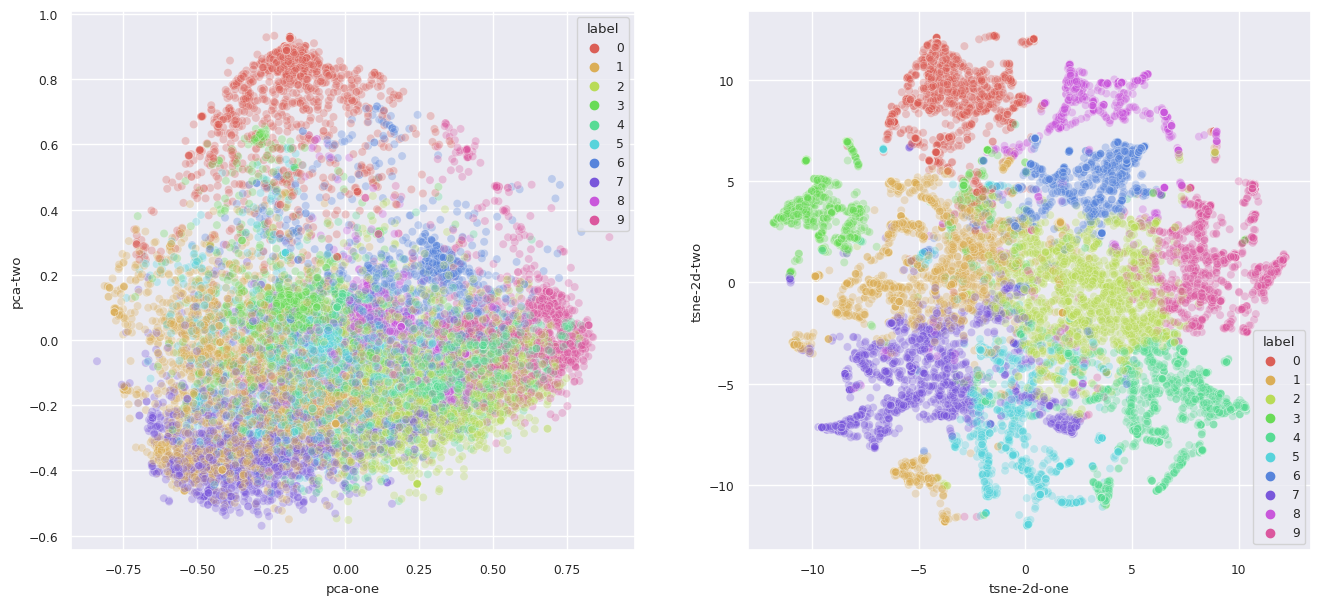}
    \caption{Clustering among all considered DAO projects}
    \label{fig:14}
    \vspace{-0.1in}
\end{figure}

\smallskip
\noindent\textbf{Clustering the Voting Contexts.}
The clustering among all considered DAO projects is investigated in \textcolor{teal}{Fig.\ref{fig:14}}. We apply K-means clustering \cite{lloyd1982least} to analyze and categorize different DAO projects based on the textual features extracted from their titles. As the titles are strings, we first preprocess the dataset by transforming the textual data into numerical representations, such as term frequency-inverse document frequency (TF-IDF) or word embeddings.

To effectively visualize and interpret the resulting clusters, we then utilize two widely-used dimensionality reduction techniques, namely Principal Component Analysis (PCA) \cite{hotelling1933analysis} and t-Distributed Stochastic Neighbor Embedding (t-SNE) \cite{van2008visualizing}. PCA is a linear technique that identifies directions of maximum variance, transforming the high-dimensional data into a one-dimensional representation (pca-one). This provides an intuitive visualization and interpretation of the resulting clusters, preserving as much of the original variance as possible.
Conversely, t-SNE is a nonlinear technique that preserves the local structure of the data, capturing complex patterns and relationships. We use t-SNE to generate a two-dimensional representation (tsne-2d-one) of the DAO project titles, enabling a detailed examination of clusters, substructures, and intricate relationships among the projects. By conducting both, we obtain a richer understanding of the underlying patterns and relationships among different DAO projects based on their titles.

As a result, we cluster the projects into 10 labels (cf. \textcolor{teal}{Fig.\ref{fig:14}}) along with brief summaries as presented below:

\smallskip
\noindent\textbf{Label 0} ${\color[rgb]{1,0.3,0.3} \blacksquare}$ \textcolor{gray}{<Protocol Upgrades and Implementations>}.
% Subtitle: Enhancing Decentralized Systems
This category focuses on proposals and discussions related to upgrades, implementations, and enhancements of decentralized protocols and platforms. Topics include network upgrades, smart contract implementations, and consensus mechanism improvements.

\smallskip
\noindent\textbf{Label 1} ${\color[rgb]{1,0.7,0.4} \blacksquare}$ \textcolor{gray}{<Governance and Decision-making>}.
% Subtitle: Steering Decentralized Organizations
This category primarily covers various proposals and discussions related to governance, management, and decision-making processes within DAOs and other decentralized organizations. Topics include voting systems, governance structure, and various aspects of administration.

\smallskip
\noindent\textbf{Label 2} ${\color[rgb]{0.85,0.95,0.05} \blacksquare}$ \textcolor{gray}{<Tokenomics, Staking, and Rewards>}.
% Subtitle: Financial Incentives in Decentralization
This category is focused on tokenomics, staking, rewards, and incentives for decentralized platforms and protocols. Discussions and proposals revolve around token distribution, staking mechanisms, yield farming, liquidity provision, and other related financial aspects.

\smallskip
\noindent\textbf{Label 3} ${\color[rgb]{0.6,0.95,0.2} \blacksquare}$ \textcolor{gray}{<Development and Technical Improvements>}.
% Subtitle: Building and Enhancing Decentralized Platforms
This category deals with discussions and proposals related to the development, improvement, and maintenance of decentralized platforms, protocols, and applications. Topics include technical improvements, bug fixes, new features, and other aspects of software development.

\smallskip
\noindent\textbf{Label 4} ${\color[rgb]{0.4,1,0.5} \blacksquare}$ \textcolor{gray}{<Marketing, Branding, and Community Building>}.
% Subtitle: Expanding Decentralized Adoption
This category covers marketing, branding, and community-building efforts within the decentralized ecosystem. Topics include community engagement, social media presence, promotional campaigns, partnerships, and collaborations to increase visibility and adoption.

\smallskip
\noindent\textbf{Label 5} ${\color[rgb]{0.55,1,0.98} \blacksquare}$ \textcolor{gray}{<Budgets, Funding, and Financial Management>}.
% Subtitle: Resource Allocation in Decentralization
This category focuses on various budgets, funding, and financial aspects related to decentralized organizations and projects. Discussions and proposals revolve around allocating resources, managing expenses, funding proposals, and other financial matters.

\smallskip
\noindent\textbf{Label 6} ${\color[rgb]{0,0.4,0.8} \blacksquare}$ \textcolor{gray}{<Project-related Requests and Resources>}.
% Subtitle: Empowering Decentralized Initiatives
This category encompasses project-related requests, including requests for resources, support, or collaboration from the community. Topics include project funding, hiring, development services, and other resources needed to move a project forward.

\smallskip
\noindent\textbf{Label 7.} ${\color[rgb]{0.4,0,0.8} \blacksquare}$ \textcolor{gray}{<Asset Management and Acquisitions>}.
% Subtitle: Digital Assets in the Decentralized World
This category deals with asset management, acquisitions, and purchases within the decentralized ecosystem. Topics include buying and selling NFTs, real estate in virtual worlds, and other digital assets, as well as decisions regarding strategic investments or acquisitions.

\smallskip
\noindent\textbf{Label 8} ${\color[rgb]{0.8,0,0.8} \blacksquare}$ \textcolor{gray}{<Contests, Competitions, and Events>}.
% Subtitle: Engaging the Decentralized Community
This category is focused on contests, competitions, programs, and events within the decentralized ecosystem. Discussions and proposals revolve around voting on the outcomes of various competitions, participating in events or programs, and other community engagement activities.

\smallskip
\noindent\textbf{Label 9} ${\color[rgb]{0.9,0,0.6} \blacksquare}$ \textcolor{gray}{<Activation and Continuation>}.
% Subtitle: Managing People and Projects in Decentralization
This category covers the activation and continuation of individuals or projects within decentralized organizations. Topics include activating new members, continuing or ending ongoing initiatives, adjusting reward structures, and other decisions related to the management of human resources and projects.

\smallskip
\noindent\textbf{Relations between Labels.} Evidence of correspondence between the label descriptions and the clustering outcomes can be observed in \textcolor{teal}{Fig.\ref{fig:14}} through select examples. 
Label 0 ${\color[rgb]{1,0.7,0.4} \blacksquare}$ adjacent to Label 3 ${\color[rgb]{0.6,0.95,0.2} \blacksquare}$ signifies the close relationship between executing protocol upgrades and specific development and technical improvements. Label 4 ${\color[rgb]{0.4,1,0.5} \blacksquare}$ encompasses marketing-related events and is closely linked with financial management represented by Label 5 ${\color[rgb]{0.55,1,0.98} \blacksquare}$ and asset management represented by Label 7 ${\color[rgb]{0.4,0,0.8} \blacksquare}$.
Label 8 ${\color[rgb]{0.8,0,0.8} \blacksquare}$, concentrating on contest events, stems from the initiatives and activation characterized by Label 9 ${\color[rgb]{0.9,0,0.6} \blacksquare}$. Simultaneously, both Label 8 ${\color[rgb]{0.8,0,0.8} \blacksquare}$ and Label 9 ${\color[rgb]{0.9,0,0.6} \blacksquare}$ intersect with Label 2 ${\color[rgb]{0.85,0.95,0.05} \blacksquare}$, highlighting that the activation and continuation of individuals or projects within DAOs necessitate extensive discussions about adopting appropriate token incentives. 
Conversely, Label 1 ${\color[rgb]{1,0.7,0.4} \blacksquare}$ and Label 2 ${\color[rgb]{0.85,0.95,0.05} \blacksquare}$, positioned at the center, validate that the governance and tokenomics components form the core of DAOs, aligning with the introduction presented in \textcolor{magenta}{Sec.\ref{sec:dao_components}}.

\begin{center}
\tcbset{
        enhanced,
      % colback=black!5!white,
        boxrule=0pt,
      %  colframe=black!75!black,
        fonttitle=\bfseries
       }
       \begin{tcolorbox}[
     %  title=Taking-home Messages,
       lifted shadow={1mm}{-2mm}{3mm}{0.1mm}{black!50!white}
       ]
      \textbf{\text{Insight-\ding{205}:}} DAOs currently exhibit a broad range of voting contexts, covering topics from budget allocations and project funding to community events and hiring decisions. This diversity showcases the potential for decentralized governance to empower communities and drive innovation across various domains. However, challenges such as voter apathy and the concentration of power among a few token holders highlight the need for more robust, inclusive, and accessible governance mechanisms that encourage broader participation and ensure a sustainable future for DAOs.
       
       \end{tcolorbox}
\end{center}
\vspace{-0.1in}

%%%%-----------------------------------
\subsection{DAO Tokens Usage}
This sector describes the usage of different \textit{DAO tokens} used in the considered DAO projects or proposals.

\begin{figure}[t]
\centering
\subfigure[Fraction of all tokens (general)]{
\resizebox{0.22\textwidth}{!}{
    \includegraphics[width=0.9\linewidth]{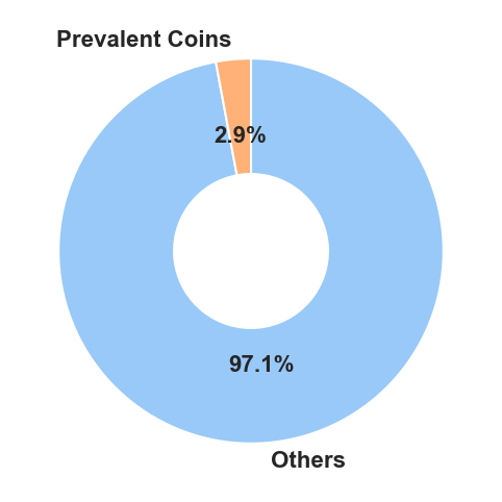}
\label{fig:2a}
}
}
%\hspace{2em}
\subfigure[Fraction of prevalent tokens]{
\resizebox{0.22\textwidth}{!}{
    \includegraphics[width=0.9\linewidth]{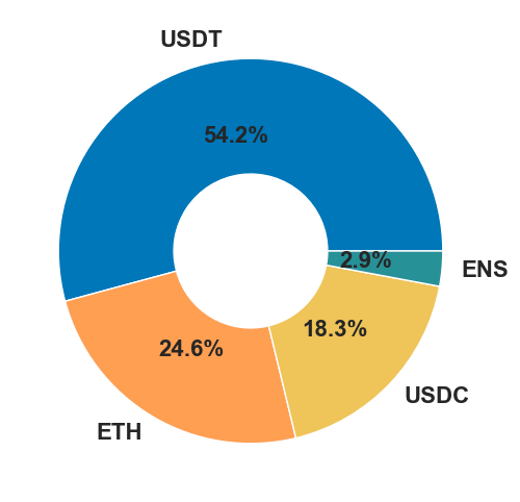}
\label{fig:2b}
}
}

\subfigure[Fraction of self-issued tokens]{
\resizebox{0.48\textwidth}{!}{
    \includegraphics[width=0.9\linewidth]{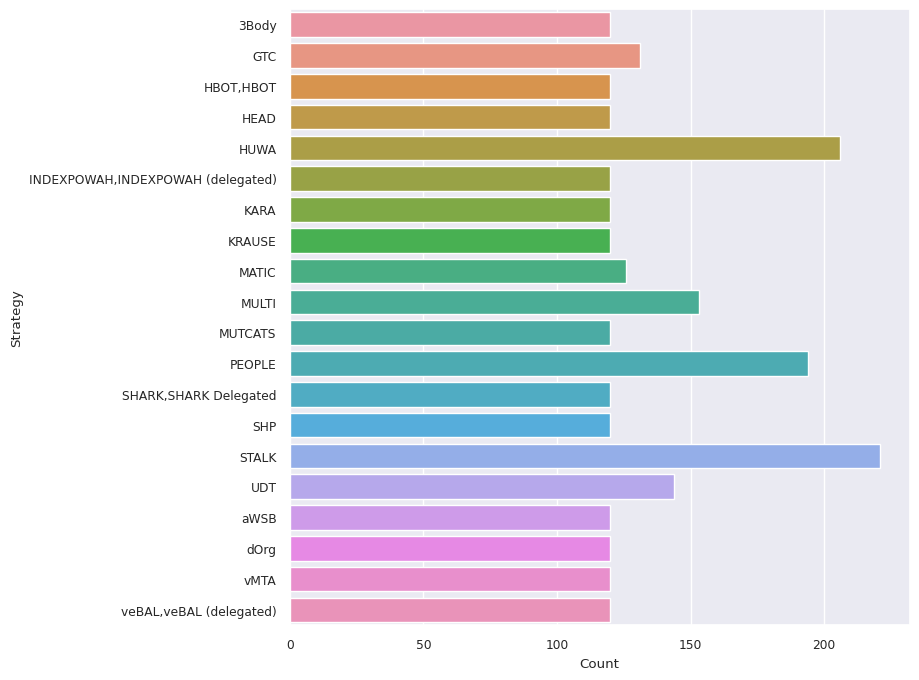}
\label{fig:2c}
}
}
\vspace{-0.1in}
\caption{Distribution of the Token Usage}
\label{fig:4:e-voting}
\vspace{-0.1in}
\end{figure}

\textcolor{teal}{Fig.\ref{fig:2a}} reveals that 97.1\% of the DAO projects use self-issued (equiv. customized) tokens or minor tokens, while only 2.9\% of the DAO projects use the mainstream tokens including USDT (54.2\%), ETH (24.6\%), USDC (18.3\%), and ENS (2.9\%), as shown in \textcolor{teal}{Fig.\ref{fig:2b}}. The results reveal a risk of the current usage of tokens in DAO spaces. The majority stays on using self-issued tokens or minor tokens which are much less stable and have much fewer merits than the prevalent tokens. Unhealthy opportunistic behaviors could be apparently realized, which is adverse to leveraging smooth and efficient governance. Across the DAOs using self-issued tokens, the top 3 are \textit{STALK}, \textit{HUWA}, and \textit{PEOPLE} whereas \textit{HUWA} is tailored specifically to internet memes compared to \textit{STALK} facilitating a fiat stablecoin protocol and \textit{PEOPLE} aiming to develop the subDAOs, as shown in \textcolor{teal}{Fig.\ref{fig:2c}}. This also implies the immaturity of the DAO community and needs further improvement. 

Another interesting observation is that most customized tokens (over 75\% among ``Others'' in \textcolor{teal}{Fig.\ref{fig:2a}}) are minted on the top of Ethereum ecosystems, which means they are intriguingly designed in forms of \textit{ERC-20} tokens that are closely relied on the development of Ethereum platforms. Similarly, the rest of the customized tokens are created on other mainstream public chains, such as BSC and Avalanche. Such situations indicate a potential threat of implicit centralization caused by oligopolistic blockchain organizations that have taken the first-mover advantages.

\begin{center}
\tcbset{
        enhanced,
      % colback=black!5!white,
        boxrule=0pt,
      %  colframe=black!75!black,
        fonttitle=\bfseries
       }
       \begin{tcolorbox}[
     %  title=Taking-home Messages,
       lifted shadow={1mm}{-2mm}{3mm}{0.1mm}{black!50!white}
       ]
      \textbf{\text{Insight-\ding{206}:}} Unhealthy opportunistic behaviors are still common in the current DAO community in the sense that the majority of the projects rather relies on self-issued tokens than the apparently more valuable and stable mainstream tokens such as USDT, ETH, etc.
       \end{tcolorbox}
\end{center}

%-----------------------------------------
\section{Discussions on Threats}
\label{sec-discuss}
%-----------------------------------------

This section highlights potential challenges. The analysis of threats is largely based on empirical evidence gathered through our study.% that DAOs may face

\smallskip
\noindent\textbf{Centralization.}
The governance in DAOs relies prominently on the possession of stakes or utility tokens. Although it is originally expected to be core to the decentralization in DAOs, highly active groups of participants are likely to accumulate major shares of tokens (investigated by our \textcolor{teal}{results Fig.\ref{fig:2a}}\&\textcolor{teal}{\ref{fig:2b}}), hence breaching the decentralization due to the concentration of e-voting power. Beyond that, it's disheartening to observe the growing centralization of various aspects within DAOs. For instance, language usage (see \textcolor{teal}{Fig.\ref{fig:12}}), voting strategy (\textcolor{teal}{Fig.\ref{fig:4}}), platform adoption (\textcolor{teal}{Fig.\ref{fig:3}}), and even storage (\textcolor{teal}{Fig.\ref{fig:11}}) all seem to be following a similar path towards centralization. This trend has been discussed in depth in our finding from \textbf{Insight-\ding{202}}. Such a phenomenon raises questions about whether we can truly achieve the promise of decentralized governance in the long run.

To avoid centralization, DAOs could accordingly prioritize diversity, decentralize decision-making, avoid concentration of assets, embrace transparency, and foster community. Having a diverse group of participants from different backgrounds and expertise can prevent power from being concentrated in the hands of a few. Decentralizing decision-making by allowing all members to participate in governance and voting, through mechanisms like quadratic voting and delegation, can prevent the decision-making process from being controlled by a small group. Avoiding the concentration of assets in a single wallet or exchange can reduce the risk of a single point of failure. Embracing transparency by making all decisions and transactions publicly visible can prevent any hidden centralization from occurring. Additionally,  fostering a sense of community among members, despite it is pretty difficult, may help ensure that everyone feels invested in the success of the organization.

\smallskip
\noindent\textbf{Disunity and Fairness.}
DAO communities come across disagreements much more often than a traditional organization does (cf. \textcolor{teal}{Fig.\ref{fig:2c}}, also mentioned in \cite{feichtinger2023hidden}\cite{sharma2023unpacking})). While this reflects the democratic nature of DAOs, it also highlights the potential for disagreements to divide the community. A disagreement can arise over a wide range of issues such as strategic direction, resource allocation, or operational procedures. If left unresolved, a disagreement can escalate and lead to the formation of factions within the community. These factions may then compete against each other for power, which can undermine the decentralized nature of the DAO.

It's important to have effective mechanisms in place to resolve disagreements in a fair and transparent manner. DAOs can consider implementing dispute resolution protocols or mediation processes to address disagreements and prevent them from dividing the community. By addressing disagreements proactively and collaboratively, DAOs can maintain their democratic and decentralized nature while avoiding factionalism and preserving their collective decision-making power. Additionally, DAO governance should dictate any progress updates for the project source code or other initiatives in a fully transparent way via public communication channels, e.g., Discord, and Slack.

\smallskip
\noindent\textbf{Legality.} It is evident that the majority of successful DAOs operate within the financial sector (cf. \textcolor{teal}{Fig.\ref{fig:1}}\&\textcolor{teal}{Fig.\ref{fig:2c}}\&\textcolor{teal}{Fig.\ref{fig:14}}, \textbf{Insight-}\ding{208}), which poses significant risks from various fronts. These risks include potential attacks from malicious actors, as well as the threat of being censored by governmental entities (e.g., more than 51\% block proposers in Ethereum 2.0 are OFACed by U.S. government,  referring to our \textbf{Insight-}\ding{207}). As a result, smaller organizations may face severe limitations on their ability to operate effectively, and in some cases, these risks could even lead to their demise.

Properly embracing legal regulation can avoid the above problems. Laws or regulations about blockchain governance need to be properly established by standardizing the structures, processes, developments, and the use of blockchain and making every component (e.g., DAO) compliant with legal regulations and ethical responsibilities~\cite{kiayias2022sok}\cite{liu2022pattern}. In particular, after \textit{The DAO} hack, DAOs started to be concerned about being legally managed with better security and protection in several countries and regions \cite{blemus2017law}.

\smallskip
\noindent\textbf{High Cost.} Running a DAO on-chain can be expensive (\textcolor{teal}{Fig.\ref{fig:4:e-voting}}), with costs varying based on factors such as the underlying blockchain platform, complexity of smart contracts, and transaction volume. These costs are incurred through gas fees paid to the network, which are collected by miners or arbitrage bots and can become expensive in US dollars. Many DAOs create their own ERC20 tokens (\textcolor{teal}{Fig.\ref{fig:2c}}) to use as governance votes, which also incurs gas fees with each action taken. Even DAOs that use stablecoins (\textcolor{teal}{Fig.\ref{fig:2b}}) for voting power still need to purchase or borrow the coins from exchanges, adding to the expenses. Additionally, fees for development, maintenance, auditing, security assessments, marketing, and community building can be difficult to quantify and are excluded. 

A reasonable way to reduce the costs of operating a DAO is to rely on off-chain or layer-two techniques that can execute most operations locally. Snapshot is an off-chain platform designed to manage DAOs and enable votes. Additionally, other off-chain tools can be found in \underline{Table~\ref{tab-summary}} to further reduce costs. By leveraging these techniques, DAO operators can minimize their reliance on costly on-chain transactions and reduce their overall expenses.

\smallskip
\noindent\textbf{Nonsense Governance Activity.} After analyzing the voting contexts (e.g., proposal titles and topics), we have found that a non-negligible proportion of governance activities are nonsensical in nature (consistent with a recent report by \cite{feichtinger2023hidden}). Our analysis reveals that a considerable number of proposals (approximately 17.7\% of all proposals, raw data of \textcolor{teal}{Fig.\ref{fig:14}}) are completely irrelevant to the project's development, and merely consist of inappropriate or offensive content such as jokes and impolite questions. We think that the current ease of proposal creation, which allows anyone to submit a proposal, has contributed to the prevalence of such nonsensical activities within the governance process.

Thus, the implementation of more stringent entry requirements for proposal creation is necessary, such as mandatory completion of a tutorial on governance principles or holding a minimum number of project tokens. By introducing such measures, we expect to see an improvement in the overall quality of proposals and a reduction in the number of frivolous or fraudulent proposals.
In addition, we recommend the establishment of a mechanism to flag and remove any proposals that violate the platform's terms of service or are deemed inappropriate by the community. This could be done through the appointment of community moderators or the development of automated systems to detect such proposals.

\smallskip
\noindent\textbf{Contract reliance.} Most of the DAOs rely prominently on the authenticity and validity of the smart contracts that offer trustless environments. This implies that the vulnerability of smart contract codes and implicit design pitfalls will pose potential threats to running DAOs.  A famous historical example caused by contract pitfalls is the huge failure of \textit{The DAO} hack due to a severe bug in its smart contract code \cite{mehar2019understanding}. TheDAO raised \$150M+ (by ETH) for building a collective investment platform. However, the project crashed shortly afterward due to a severe bug in its smart contract code. As a result, a considerable amount of assets was siphoned off and a disruptive hard fork happened that significantly affected the entire Ethereum blockchain till now \cite{thedao}. Attacks like flash loans \cite{qin2021attacking} in DeFi protocols that exploit the time interval of block confirmation can also undermine the sustainability of DAO communities.

To avoid a repeat of this fiasco and stabilize the monetization mechanisms for long-term growth, DAO communities should spare efforts to establish security protocols for auditing code and develop improved tools or supportive infrastructure. Also, they need a robust marketing and product design department and develop an effective product and content strategy associated with the concepts and principles of each DAO project. Meanwhile, well-organized and consistent communication plans are also crucial to attract attention from a broader public to establish loyalty in a decentralized context.

%-----------------------------------------
\section{Further Actions}
%\section{Towards Building A Better DAO}
\label{sec-finding}
%-----------------------------------------

In this section, we continue the discussions of previous solutions and conduct a more detailed analysis of each category.

\subsection{On Projects}

Each project involves both competition and cooperation, and we will discuss them from these two perspectives.

\smallskip
\noindent\textbf{DAO-2-DAO Collaboration.}
The interaction and collaboration between different DAOs matter, spawning the design of decentralized negotiation protocols~\cite{dao2dao}. Decentralized governance in DAOs relies prominently on negotiation protocols since a unique set of parameters for forming consensus with its community, infrastructure, and use case are defined by each DAO based on a general formalization of each component such as the proposal format. This allows the evaluation of each proposal to be routinized and improves the efficiency of interactions such as joint ventures, token swaps, and distributed monetary policy. Moreover, a crafted formalization can bring DAO-2-DAO collaboration towards an inter-organizational framework by enabling proposals of different DAOs to be extended to serve a large variety of complicated contracts.

\smallskip
\noindent\textbf{Learn on SubDAOs -- Management/Competition.}
DAO management has been evolving to feature a tree structure indicating the hierarchy of different DAOs where ones might belong to the others. There will be new groups of members that operate independently of the group's inception as DAOs grow. New divisions, teams, focus, and ideas will be brought into the community. Rather than trying to house all that activity under one roof, SubDAOs are an emerging approach for different working groups to create their own foundation and ownership structure~\cite{subDAOs}. All the SubDAOs tie value back to the originating entity. At the same time, one thing to be noted is the competition among different subDAOs within the same domain. Multiple DAO participants will compete for one goal set by its superior nodes. Balanced-off games among subDAOs should be further considered for such scenarios.

\subsection{On Infrastructure}

n addition to guaranteeing the secure operation of the core blockchain, a well-developed infrastructure and a range of useful applications are crucial for promoting the widespread adoption of DAOs.

\smallskip
\noindent\textbf{DAO Stacks and Tools.} As a generic term, the DAO space has included a variety of projects that cover many components and fields. We could sketch a relatively clear picture by learning from its ``stack" (Daostack \cite{daostack}). The foundation is the basic and backend software modules such as voting mechanisms (as discussed before) for decentralized governance. On top of it, a library layer used to build models for back ends is established (e.g., Arc.js \cite{arcjs}). Also, a caching layer is needed for collecting and structuring data (e.g., The Graph \cite{thegraph}). On the top, the application layer is designed for DAO users to deploy or participate in DAOs (Aragon \cite{aragon}). In addition, a variety of widely used coordination tools, such as Twitter, Discord, and Github, play a role in supporting and facilitating DAO games from an external perspective.

\smallskip
\noindent\textbf{Applications via DAO.} 
DAOs have been considered as one of the biggest innovations in the Blockchain ecosystems \cite{chen2018blockchain}. Therein, crowdfunding is one of the prime applications where DAO plays a vital role. For instance, ConstitutionDAO successfully pulled together \$47 million worth of ether in a week to try to buy a first-edition copy of the U.S. Constitution at a Sotheby's auction \cite{Constitu10:online}. Besides, DAO has been involved in democratizing the Metaverse ecosystem by offering contributions to decentralized infrastructure \cite{gadekallu2022blockchain}. In addition, the paradigm of DAO paradigm is also applied by NFT-based investment projects in order to create and confirm shared ownership of assets. The emergence of a new generation of Dapps via DAO in various sectors, e.g., supply chain, finance accounting, IoT, healthcare, and transportation \cite{el2020overview} has demonstrated the innovation and the need for DAO in current technology trends. Especially, DAO is also investigated that it could be promising for E-government systems in improving the efficiency and transparency of government operations \cite{diallo2018egov}.

\subsection{On Voting Strategies}

Two key questions regarding voting are: \textit{how to cast a vote} and \textit{how the outcome of the vote impacts decisions}.

\smallskip
\noindent\textbf{Voting Routes.} Voting could be conducted through both on-chain or off-chain. The on-chain voting service, such as Tally \cite{tally}, has to introduce the \textit{time-lock} mechanism to provide the polling period. Implementing such a mechanism typically relies on the usage of smart contracts. Tally's voting contains two types of smart contracts: a \textit{token} contract and a \textit{governor}  contract. Meanwhile, the multi-sig wallet (e.g., Gnosis Safe \cite{gnosissafe}) is necessary for managing the deployed assets. However, on-chain voting confronts the disadvantages of \textit{costly} and \textit{delayed confirmation}, significantly decreasing the willingness of participation of users. In contrast, Snapshot is an off-chain voting tool that removes the expensive consumption of on-chain interactive operations. The number of created DAO spaces in two platforms indicates that users have much more willing to participate and vote in a gas-free platform. 

\smallskip
\noindent\textbf{Strategy Design.}  The design of voting strategies in DAOs plays a crucial role in ensuring effective decision-making and fostering user participation. These strategies should strike a balance between security, efficiency, and inclusiveness, accommodating various voting tools, applications, and regulatory requirements.
On-chain and off-chain voting methods can be combined to create hybrid strategies, leveraging the strengths of each approach. For instance, off-chain voting through tools like Snapshot can be employed for preliminary or less critical decisions, allowing for a more agile and gas-free voting process. On the other hand, on-chain voting, such as the Tally mechanism, can be reserved for more critical decisions, where the security and immutability provided by blockchain technology are essential.
Another dimension to consider in voting strategy design is the DAO-to-DAO voting mechanism, where one DAO can participate in the decision-making process of another DAO. This can promote cross-DAO collaboration and resource sharing, fostering synergies within the decentralized ecosystem. Voting in SubDAOs can also be utilized to facilitate the delegation of decision-making power to specialized groups, enabling efficiency in governance.

\subsection{On Tokenization}\label{subsec-tokenization}

Tokenization forms the foundation of the blockchain economy and incentive mechanisms. Achieving sustainability and a healthy Web3 ecosystem requires building on tokenization.

\smallskip
\noindent\textbf{Healthy \textcolor{gray}{v.s.} Unhealthy Tokenization.}  
A healthy tokenization distribution enables fairness to people who are involved in the DAO projects. It means that anyone who is purchasing the token competes on the same terms and is subjected to the same token sales policies. Besides, reasonable token usage of DAOs significantly impacts project-controlled liquidity\cite{warren20170x}. As discussed in the previous section, the imbalance between self-issued tokens and mainstream tokens raises the risk of manipulation in the market. The equality of token usage logic has significant implications. For example, small market cap projects launched at cheap initial prices usually face the potential abuse of whale and team purchases. Meanwhile, the larger the market capitalization which is managed by DAOs, the less likely the chances that the whale and inside team can purchase tokens on the market. Hence, the balance between the mainstream tokens and self-issued tokens not only reduces the possibility of market manipulation but also provides space and finances for the founding team to develop the DAO projects.

\smallskip
\noindent\textbf{Governance via Tokenization.} Effective governance is crucial for aligning the interests of various stakeholders and ensuring the stability of the ecosystem. This is particularly important in DAOs where incentivizing responsible behavior can be challenging. Tokenization can be used to provide \textit{monetary incentives} to all parties involved, including the project team, application providers, node operators, blockchain users, and even regulators. However, ensuring fairness in the distribution of incentives is equally important. A well-designed governance structure should incorporate game theory to encourage diverse stakeholder participation and representation. Additionally, transparent distribution of on-chain and off-chain incentives can help build trust and cooperation among stakeholders toward achieving common goals. At a higher level, maintaining a balance between mainstream tokens and self-issued tokens can reduce the risk of market manipulation. Building better governance through rational incentives and transparent mechanisms can lead to a Schelling point~\cite{LIU2022102090}, where desirable behaviors are encouraged, and fairness is maintained.

\smallskip
% Ref-1: https://blog.curvelabs.eu/d2d-towards-decentralized-negotiation-protocols-e37d164e91e6

% Ref: https://coopahtroopa.mirror.xyz/7bfK9st2mvhxlla4XKotRjetq5-YhaiwqRwS8DhkD-o

%-----------------------------------------
\section{Related Work}
\label{sec-rw}
%-----------------------------------------
This section covers three dimensions of DAO progress: the evolution of several major DAOs in the industry, formative research on DAOs, and related work on Web3 governance.

\smallskip
\noindent\textbf{DAO Evolution.} We callback several milestones in DAO's history. The first DAO, known as \textit{The DAO} \cite{thedaostory}, was established on Ethereum in 2016, marking the beginning of DAOs on blockchains. Unfortunately, the project was hacked, ultimately leading to a hard fork of the Ethereum blockchain \cite{thedao}\cite{mehar2019understanding}. After this setback, DAOs regained popularity with the emergence of MakerDAO \cite{makerdao} in 2018. The project introduced an on-chain governance system to produce a deposited stablecoin protocol (a.k.a. DAI). Then, in 2020, a surge of decentralized finance (DeFi) protocols \cite{werner2021sok}, known as the \textit{DeFi summer}, propelled DAOs to new heights. These protocols are built on top of various blockchain platforms, such as Ethereum, BSC, and Avalanche, and enable decentralized finance services such as DEXs (Uniswap, dYdX), lending (Compound, Aave), yield aggregators (Convex), and staking (Lido), among others. Till now, DAOs embraced the concept of Web3 \cite{wang2022exploring}, and their development became intertwined with the surrounding components that make up the decentralized web. This includes elements such as wallets, smart contracts, various blockchain platforms, and even regulations \cite{daocomplexities}.

\smallskip
\noindent\textbf{Formative DAO Studies.} Liu et al.~\cite{liu2021technology} provide an overview of early DAOs by explaining the definitions, and preliminaries and giving a simple taxonomy. Daian shows a series of DAO attack analyses~\cite{daoexploit}\cite{daowake} from the technical level by diving into the source code. They point out the reasons for recursive send vulnerabilities in Ethereum that cause a monetary loss (\$150M). Robin et al. \cite{fritsch2022analyzing} have investigated three DAO projects (Compound, Uniswap, and ENS) by empirically analyzing their voting powers and discussing governance. Later, Daian et al.~\cite{darkdao} propose a potential attack form called \textit{Dark DAO}, which means a group of members form a decentralized cartel and can opaquely manipulate (e.g., buy) on-chain votes. Yu et al. \cite{yu2022leveraging} provide a quick review of existing DAO literature and deliver their responses by statistically reviewed papers. Feichtinger et al.~\cite{feichtinger2023hidden} conduct an empirical study on 21 DAOs to explore the hidden problems, including high centralization, monetary costs, and pointless activities. Sharma et al.~\cite{sharma2023unpacking} have developed their 
research on 10 diverse DAOs to examine their degree of decentralization and autonomy. Besides, many researchers and organizations also put their focus on DAO by generating in-time reports~\cite{daoreport}\cite{daosam} and online posts \cite{daocomplexities}. Instead of concentrating on single attacks or individual projects, we provide an in-depth analysis of DAO projects resident on the Snapshot platform. 

\smallskip
\noindent\textbf{Governance in Web3.} Governance is the cornerstone of DAOs, but research in this area remains vague. DAOs are rooted in blockchain and heavily influenced by on-chain tokens, which means they are affected by the underlying technology as well as cryptocurrency regulations. For the former, Kiayias et al.~\cite{kiayias2022sok} have explored the governance process and properties within the blockchain context. They use a first principles approach to derive seven fundamental properties and apply them to evaluate existing projects. Wang et al.~\cite{wang2022exploring} have partially incorporated this notion into Web3, emphasizing the importance of DAO governance in Web3. Liu et al.~\cite{liu2022systematic} have presented a systematic review of primary studies on governance and provide a qualitative and quantitative synthesis. Regarding the latter, governance primarily comes from governments of different countries, even though blockchain is designed to be decentralized. However, it can still be easily regulated by increasingly centralized miners or validators. Ethereum (after the Merge) PoS validators who rely on MEV-boots services may be monitored and regulated by the government (OFAC-compliant). Additionally, pure cryptocurrency companies can also be governed. A noteworthy example is the US sanctions against Tornadocash~\cite{tornadocashsanction}.

%-----------------------------------------
\section{Conclusion}
\label{sec-conclu}
%-----------------------------------------
%\smallskip
%\noindent\textbf{\textcolor{black}{Conclusion.}} 

In this paper, we empirically studied the ground truth of the wild DAO projects collected from the largest off-chain voting platform, \textit{Snapshot}. We conducted comprehensive data collection and analysis, and shed insights on the design principles of the prevalent DAOs. Our empirical results discover a series of facts hidden from the light, covering its unfair distributions on such as participant regions or language usage, as well as revealing potential threats of centralization and contract reliance. We also showed our minds ways of building a better DAO from different aspects. This work, at least in our opinion, delineates a clear view of current DAOs.

%============================================
\bibliographystyle{unsrt}
%\bibliography{bib}
\bibliography{bib(aft)}

\begin{thebibliography}{10}

\bibitem{wang2022exploring}
Qin Wang, Rujia Li, et~al.
\newblock Exploring {Web3} from the view of blockchain.
\newblock {\em arXiv preprint arXiv:2206.08821}, 2022.

\bibitem{liu2021technology}
Lu~Liu, Sicong Zhou, et~al.
\newblock From technology to society: An overview of blockchain-based {DAO}.
\newblock {\em IEEE Open Journal of the Computer Society}, 2:204--215, 2021.

\bibitem{yu2022leveraging}
Guangsheng Yu, Qin Wang, Tingting Bi, Shiping Chen, and Sherry Xu.
\newblock Leveraging architectural approaches in {Web3} applications--a {DAO}
  perspective focused.
\newblock {\em IEEE International Conference on Blockchain and Cryptocurrency
  Workshop on Cryptocurrency Exchanges (CryptoEx@ICBC)}, 2023.

\bibitem{faqir2021comparative}
Youssef Faqir-Rhazoui et~al.
\newblock {A Comparative Analysis of the Platforms for Decentralized Autonomous
  Organizations in the Ethereum Blockchain}.
\newblock {\em Journal of Internet Services and Applications}, 12/1(1):1--20,
  2021.

\bibitem{wang2019decentralized}
Shuai Wang, Wenwen Ding, Juanjuan Li, and other.
\newblock {Decentralized Autonomous Organizations: Concept, Model, and
  Applications}.
\newblock {\em IEEE Transactions on Computational Social Systems},
  6(5):870--878, 2019.

\bibitem{bischof2022longevity}
Evelyne Bischof, Alex Botezatu, Sergey Jakimov, Ilya Suharenko, Alexey
  Ostrovski, Andrey Verbitsky, Yury Yanovich, Alex Zhavoronkov, and Garri
  Zmudze.
\newblock Longevity foundation: Perspective on decentralized autonomous
  organization for special-purpose financing.
\newblock {\em IEEE Access}, 10:33048--33058, 2022.

\bibitem{mehdi2022data}
Ibrahim Mehdi et~al.
\newblock {Data Centric {DAO}: When blockchain reigns over the Cloud}.
\newblock In {\em IEEE International IOT, Electronics and Mechatronics
  Conference (IEMTRONICS)}, pages 1--7. IEEE, 2022.

\bibitem{feichtinger2023hidden}
Rainer Feichtinger, Robin Fritsch, et~al.
\newblock The hidden shortcomings of {DAOs}--an empirical study of on-chain
  governance.
\newblock {\em arXiv preprint arXiv:2302.12125}, 2023.

\bibitem{fritsch2022analyzing}
Robin Fritsch, Marino M{\"u}ller, et~al.
\newblock Analyzing voting power in decentralized governance: Who controls
  {DAOs}?
\newblock {\em arXiv preprint arXiv:2204.01176}, 2022.

\bibitem{sharma2023unpacking}
Tanusree Sharma, Yujin Kwon, Kornrapat Pongmala, Henry Wang, Andrew Miller,
  Dawn Song, and Yang Wang.
\newblock Unpacking how decentralized autonomous organizations (daos) work in
  practice.
\newblock {\em arXiv preprint arXiv:2304.09822}, 2023.

\bibitem{snapshot}
Snapshot.
\newblock {\em Accessible at \url{https://snapshot.org/\#/}}, 2022.

\bibitem{wood2014ethereum}
Gavin Wood et~al.
\newblock Ethereum: A secure decentralised generalised transaction ledger.
\newblock {\em Ethereum Yellow Paper}, 151(2014):1--32, 2014.

\bibitem{avalanche}
Avalanche network.
\newblock {\em Accessible at \url{https://www.avax.network/}}, 2022.

\bibitem{bsc}
Binance smart chain.
\newblock {\em Accessible at \url{https://www.bnbchain.org/en}}, 2022.

\bibitem{polygon}
Polygon network.
\newblock {\em Accessible at \url{https://polygon.technology/}}, 2022.

\bibitem{solana}
Solana network.
\newblock {\em Accessible at \url{https://solana.com/}}, 2022.

\bibitem{li2021offline}
Rujia Li et~al.
\newblock An offline delegatable cryptocurrency system.
\newblock In {\em IEEE International Conference on Blockchain and
  Cryptocurrency (ICBC)}, pages 1--9. IEEE, 2021.

\bibitem{tolmach2021survey}
Palina Tolmach, Yi~Li, et~al.
\newblock A survey of smart contract formal specification and verification.
\newblock {\em ACM Computing Surveys (CSUR)}, 54(7):1--38, 2021.

\bibitem{w3cdid}
Reed Drummond, Sporny Manu, Sabadello Markus, Longley Dave, and Allen
  Christopher.
\newblock Decentralized identifiers (\textrm{DID}s) v1.0: Core architecture,
  data model, and representations.
\newblock {\em Accessible at \url{https://www.w3.org/TR/did-core/}}, 2021.

\bibitem{ens}
Ethereum name service.
\newblock {\em Accessible at \url{https://ens.domains/}}, 2022.

\bibitem{kiayias2022sok}
Aggelos Kiayias and Philip Lazos.
\newblock {SoK}: Blockchain governance.
\newblock {\em Proceedings of the ACM Conference on Advances in Financial
  Technologies (AFT)}, 2022.

\bibitem{sybil}
John~R. Douceur.
\newblock The sybil attack.
\newblock In Peter Druschel, Frans Kaashoek, and Antony Rowstron, editors, {\em
  Peer-to-Peer Systems}, pages 251--260, Berlin, Heidelberg, 2002. Springer
  Berlin Heidelberg.

\bibitem{cortier2016sok}
V{\'e}ronique Cortier, David Galindo, Ralf K{\"u}sters, Johannes M{\"u}ller,
  and Tomasz Truderung.
\newblock {SoK}: Verifiability notions for e-voting protocols.
\newblock In {\em IEEE Symposium on Security and Privacy (SP)}, pages 779--798.
  IEEE, 2016.

\bibitem{eip20}
Fabian Vogelsteller and Vitalik Buterin.
\newblock {EIP}-20: Token standard.
\newblock {\em Ethereum Improvement Proposals, accessible at
  \url{https://eips.ethereum.org/EIPS/eip-20}}, 2015.

\bibitem{eip721}
William Entriken, Dieter Shirley, Jacob Evans, and Nastassia Sachs.
\newblock {EIP}-721: Non-fungible token standard.
\newblock {\em Ethereum Improvement Proposals, accessible at
  \url{https://eips.ethereum.org/EIPS/eip-721}}, 2018.

\bibitem{eip1155}
Radomski Witek, Cooke Andrew, Castonguay Philippe, Therien James, Binet Eric,
  and Sandford Ronan.
\newblock {EIP}-1155: Multi token standard.
\newblock {\em Ethereum Improvement Proposals, accessible at
  \url{https://eips.ethereum.org/EIPS/eip-1155}}, 2018.

\bibitem{wang2021non}
Qin Wang, Rujia Li, et~al.
\newblock Non-fungible token (nft): Overview, evaluation, opportunities and
  challenges.
\newblock {\em arXiv preprint arXiv:2105.07447}, 2021.

\bibitem{eip4824}
Tan Joshua, Patka Isaac, Gershtein Ido, Eithcowich Eyal, et~al.
\newblock Common interfaces for daos.
\newblock {\em \url{https://eips.ethereum.org/EIPS/eip-4824}}, 2022.

\bibitem{eip1202}
Zhou Zainan, Evan, and Xu~Yin.
\newblock {EIP}-1202: Voting interface [draft].
\newblock {\em Accessible at \url{https://eips.ethereum.org/EIPS/eip-1202}},
  2018.

\bibitem{eip779}
Detrio Casey.
\newblock {EIP}-779: Hardfork meta: {DAO} fork.
\newblock {\em Accessible at \url{https://eips.ethereum.org/EIPS/eip-779}},
  2017.

\bibitem{daoreport}
Slavin Aiden et~al.
\newblock World economic forum: Decentralized autonomous organizations: Beyond
  the hype.
\newblock {\em
  \url{https://www3.weforum.org/docs/WEF_Decentralized_Autonomous_Organizations_Beyond_the_Hype_2022.pdf}},
  2022.

\bibitem{benet2014ipfs}
Juan Benet.
\newblock {IPFS}-content addressed, versioned, {P2P} file system.
\newblock {\em arXiv preprint arXiv:1407.3561}, 2014.

\bibitem{pareto}
Wiki.
\newblock The pareto principle.
\newblock {\em \url{https://www.wikiwand.com/en/Pareto_principle}}, 2022.

\bibitem{pareto1964cours}
Vilfredo Pareto.
\newblock {\em Cours d'{\'e}conomie politique}, volume~1.
\newblock Librairie Droz, 1964.

\bibitem{ipfscid}
Ipfs doc: Content addressing and cids.
\newblock {\em Accessible at
  \url{https://docs.ipfs.tech/concepts/content-addressing/#cid-conversion}},
  2022.

\bibitem{matthew}
Wiki.
\newblock Matthew effect.
\newblock {\em \url{https://www.wikiwand.com/en/Matthew_effect}}, 2022.

\bibitem{lloyd1982least}
Stuart Lloyd.
\newblock Least squares quantization in pcm.
\newblock {\em IEEE Transactions on Information Theory (TIT)}, 28(2):129--137,
  1982.

\bibitem{hotelling1933analysis}
Harold Hotelling.
\newblock Analysis of a complex of statistical variables into principal
  components.
\newblock {\em Journal of Educational Psychology}, 24(6):417, 1933.

\bibitem{van2008visualizing}
Laurens Van~der Maaten and Geoffrey Hinton.
\newblock Visualizing data using t-sne.
\newblock {\em Journal of Machine Learning Research}, 9(11), 2008.

\bibitem{liu2022pattern}
Yue Liu, Qinghua Lu, Guangsheng Yu, Hye-Young Paik, Harsha Perera, and Liming
  Zhu.
\newblock A pattern language for blockchain governance.
\newblock In {\em Proceedings of the 27th European Conference on Pattern
  Languages of Programs}, pages 1--16, 2022.

\bibitem{blemus2017law}
St{\'e}phane Blemus.
\newblock Law and blockchain: A legal perspective on current regulatory trends
  worldwide.
\newblock {\em Revue Trimestrielle de Droit Financier (Corporate Finance and
  Capital Markets Law Review) RTDF}, (4-2017), 2017.

\bibitem{mehar2019understanding}
Muhammad~Izhar Mehar, Charles~Louis Shier, Alana Giambattista, Elgar Gong,
  Gabrielle Fletcher, Ryan Sanayhie, Henry~M Kim, and Marek Laskowski.
\newblock Understanding a revolutionary and flawed grand experiment in
  blockchain: the {DAO} attack.
\newblock {\em Journal of Cases on Information Technology (JCIT)},
  21(1):19--32, 2019.

\bibitem{thedao}
OLEKSII KONASHEVYCH.
\newblock Takeaways: 5 years after the {DAO} crisis and ethereum hard fork.
\newblock {\em Accessible at
  \url{https://cointelegraph.com/news/takeaways-5-years-after-the-dao-crisis-and-ethereum-hard-fork}},
  2021.

\bibitem{qin2021attacking}
Kaihua Qin, Liyi Zhou, Benjamin Livshits, and Arthur Gervais.
\newblock Attacking the {DeFi} ecosystem with flash loans for fun and profit.
\newblock In {\em International Conference on Financial Cryptography and Data
  Security (FC)}, pages 3--32. Springer, 2021.

\bibitem{dao2dao}
Cem~F Dagdelen.
\newblock {D2D}: Towards decentralized negotiation protocols.
\newblock {\em
  \url{https://blog.curvelabs.eu/d2d-towards-decentralized-negotiation-protocols-e37d164e91e6}},
  2021.

\bibitem{subDAOs}
Coopahtroopa.
\newblock How to subdao.
\newblock {\em
  \url{https://coopahtroopa.mirror.xyz/7bfK9st2mvhxlla4XKotRjetq5-YhaiwqRwS8DhkD-o}},
  2021.

\bibitem{daostack}
{DaoStack}: An operating system for collective intelligence.
\newblock {\em Accessible at \url{https://medium.com/daostack}}, 2022.

\bibitem{arcjs}
Arc.js: Daostack javascript client.
\newblock {\em \url{https://daostack.github.io/arc.js/}}, 2022.

\bibitem{thegraph}
The {Graph}: {APIs} for a decentralized future.
\newblock {\em \url{https://thegraph.com/en/}}, 2022.

\bibitem{aragon}
Aragon platform.
\newblock {\em Accessible at \url{https://aragon.org/}}, 2022.

\bibitem{chen2018blockchain}
Yan Chen.
\newblock Blockchain tokens and the potential democratization of
  entrepreneurship and innovation.
\newblock {\em Business horizons}, 61(4):567--575, 2018.

\bibitem{Constitu10:online}
Constitutiondao crypto investors lose bid to buy constitution copy.
\newblock
  \url{https://www.cnbc.com/2021/11/18/constitutiondao-crypto-investors-lose-bid-to-buy-constitution-copy.html}.
\newblock (Accessed on 11/26/2022).

\bibitem{gadekallu2022blockchain}
Thippa~Reddy Gadekallu et~al.
\newblock Blockchain for the metaverse: A review.
\newblock {\em arXiv preprint arXiv:2203.09738}, 2022.

\bibitem{el2020overview}
Youssef El~Faqir, Javier Arroyo, and Samer Hassan.
\newblock An overview of decentralized autonomous organizations on the
  blockchain.
\newblock In {\em Proceedings of the 16th international symposium on open
  collaboration}, pages 1--8, 2020.

\bibitem{diallo2018egov}
Nour Diallo, Weidong Shi, Lei Xu, Zhimin Gao, Lin Chen, Yang Lu, Nolan Shah,
  Larry Carranco, et~al.
\newblock {eGov-DAO}: A better government using blockchain based decentralized
  autonomous organization.
\newblock In {\em International Conference on eDemocracy \& eGovernment
  (ICEDEG)}, pages 166--171. IEEE, 2018.

\bibitem{tally}
Tally.
\newblock {\em Accessible at \url{https://www.tally.xyz/}}, 2022.

\bibitem{gnosissafe}
Gnosis-safe: Trusted platform to manage digital assets on ethereum.
\newblock {\em Accessible at \url{https://gnosis-safe.io/}}, 2022.

\bibitem{warren20170x}
Will Warren et~al.
\newblock 0x: An open protocol for decentralized exchange on the ethereum
  blockchain.
\newblock {\em \url{https://github.com/0xProject/whitepaper}}, 2017.

\bibitem{LIU2022102090}
Yue Liu, Qinghua Lu, Guangsheng Yu, Hye-Young Paik, and Liming Zhu.
\newblock Defining blockchain governance principles: A comprehensive framework.
\newblock {\em Information Systems}, 109:102090, 2022.

\bibitem{thedaostory}
Samuel Falkon.
\newblock The story of the {DAO} — its history and consequences.
\newblock {\em
  \url{https://medium.com/swlh/the-story-of-the-dao-its-history-and-consequences-71e6a8a551ee}},
  2017.

\bibitem{makerdao}
MakerDAO.
\newblock Makerdao foundation proposal v2.
\newblock {\em
  \url{https://medium.com/@MakerDAO/foundation-proposal-v2-f10d8ee5fe8c}},
  2018.

\bibitem{werner2021sok}
Sam~M Werner, Daniel Perez, Lewis Gudgeon, Ariah Klages-Mundt, Dominik Harz,
  and William~J Knottenbelt.
\newblock {SoK}: Decentralized finance ({DeFi}).
\newblock {\em International Conference on Financial Cryptography and Data
  Security (FC)}, 2022.

\bibitem{daocomplexities}
Chainlink.
\newblock Daos and the complexities of web3 governance.
\newblock {\em Accessible at \url{https://blog.chain.link/daos/}}, 2022.

\bibitem{daoexploit}
Daian Philip.
\newblock Analysis of the {DAO} exploit.
\newblock {\em Accessible at
  \url{https://hackingdistributed.com/2016/06/18/analysis-of-the-dao-exploit/}},
  2016.

\bibitem{daowake}
Daian Philip.
\newblock Chasing the {DAO} attacker’s wake.
\newblock {\em Accessible at
  \url{https://pdaian.com/blog/chasing-the-dao-attackers-wake/}}, 2016.

\bibitem{darkdao}
Daian Philip, Kell Tyler, Miers Ian, and Juels Ari.
\newblock On-chain vote buying and the rise of dark daos.
\newblock {\em Accessible at
  \url{https://hackingdistributed.com/2018/07/02/on-chain-vote-buying/}}, 2018.

\bibitem{daosam}
Hassan Samer and Filippi Primavera.
\newblock Decentralized autonomous organization.
\newblock {\em Internet Policy Review,
  \url{https://policyreview.info/pdf/policyreview-2021-2-1556.pdf}}.

\bibitem{liu2022systematic}
Yue Liu, Qinghua Lu, Liming Zhu, Hye-Young Paik, et~al.
\newblock A systematic literature review on blockchain governance.
\newblock {\em Journal of Systems and Software}, 2022.

\bibitem{tornadocashsanction}
{U.S.} treasury sanctions notorious virtual currency mixer tornado cash.
\newblock {\em accessible
  \url{https://home.treasury.gov/news/press-releases/jy0916}}, 2022.

\end{thebibliography}
%============================================

\begin{table*}[!hbt]
\centering
\caption{Mainstream DAOs \& Tools \& DataFeeds}\label{tab-summary}
  \begin{threeparttable}  
\resizebox{0.99\linewidth}{!}{
\begin{tabular}{ll|ccc|ccc|ccccc|cccc}
\toprule

& \multicolumn{1}{c}{ \multirow{2}{*}{\textit{\textbf{DAOs}}} } &  \multicolumn{3}{c}{\textit{\textbf{Operational Features}}} & \multicolumn{3}{c}{\textit{\textbf{Functionalities}}} & \multicolumn{5}{c}{\textit{\textbf{Non-Functionalities}}} & \multicolumn{4}{c}{\textbf{\textit{Market Performance} [May 2023]}} \\

\cmidrule{3-17}

& 
\multicolumn{1}{c}{} & \multicolumn{1}{c}{\textit{\textbf{Network}}} &  \multicolumn{1}{c}{\textit{\textbf{Protocol/Field}}} &  
\multicolumn{1}{c}{\textit{\textbf{\makecell{Token}}}} &  
\multicolumn{1}{l}{\rotatebox{90}{\textit{\textbf{Create}}}} & 
\multicolumn{1}{l}{\rotatebox{90}{\textit{\textbf{Propose}}}} & 
\multicolumn{1}{l}{\rotatebox{90}{\textit{\textbf{Vote}}}} &
\multicolumn{1}{l}{\rotatebox{90}{\textit{\textbf{Permissionless}}}}  &
\multicolumn{1}{l}{\rotatebox{90}{\textit{\textbf{Transparency}}}}  &
\multicolumn{1}{l}{\rotatebox{90}{\textit{\textbf{Anti-censorship}}}} &
\multicolumn{1}{l}{\rotatebox{90}{\textit{\textbf{Interoperability}}}}  &
\multicolumn{1}{l}{\rotatebox{90}{\textit{\textbf{Incentive}}}} &
\multicolumn{1}{c}{\rotatebox{00}{\textit{\textbf{\makecell{Treasury (USD)}}}}} &
\multicolumn{1}{c}{\rotatebox{00}{\textit{\textbf{Holders}}}} &
\multicolumn{1}{c}{\rotatebox{00}{\textit{\textbf{Proposals}}}} &
\multicolumn{1}{c}{\rotatebox{00}{\textit{\textbf{Votes}}}}  
\\

\midrule

\multirow{40}{*}{\rotatebox{90}{\textit{\textbf{Projects}}}} & Uniswap & Ethereum & DeFi (DEX) & UNI  & \cmark & \cmark & \cmark  &  \cmark & \cmark & \xmark &  \xmark  & \cmark  & 2.7B & 363k & 124 & 203.8k  \\ 

& BitDAO & Ethereum & DeFi (DEX) & BIT  & \cmark & \cmark & \cmark  &  \cmark & \cmark & \xmark &   \xmark  & \cmark & 2.7B & 18.5k & 23 & 4.8k   \\ 

& ENS & Ethereum & Name Service  & ENS  & \cmark & \cmark & \cmark  &  \cmark & \cmark & \xmark &   \xmark  & \xmark & 1.1B & 64.2k & 60 & 111.8k    \\ 

& Gnosis & Ethereum & DeFi (DEX) & GNO  & \cmark & \cmark & \cmark  &  \cmark & \cmark & \xmark & \xmark  & \cmark & 1B & 363k & 124 & 203.8k    \\ 

& dYdX & Ethereum & DeFi (Lending) &  DYDX & \cmark & \cmark & \cmark  &  \cmark & \cmark & \xmark & \xmark & \cmark & 903.5M & 36.4k & 26 & 11.1k   \\ 

& Stargate.Fin & Ethereum & Service & STG  &  \cmark &  \cmark & \cmark  &  \cmark & \cmark & n/a &   \xmark  & \cmark & 374.8M & 26.8k & 47 &  2.2M\\  

& Lido & Ethereum & DeFi (Lending) & LDO  & \cmark & \cmark & \cmark  &  \cmark & \cmark & \xmark &   \xmark  & \cmark & 352.6M & 33.2k & 128 & 42.3k \\ 

& Polkadot & Substrate & Service & DOT  & \cmark & \cmark & \cmark  &  \cmark & \cmark & n/a & \xmark  & \xmark & 280.4M & 1.3M & 363 &  2.17k\\ 

& Frax.Fin & Ethereum & Stablecoin & FXS  & \cmark & \cmark & \cmark  &  \cmark & \cmark & \xmark &  \xmark  & \xmark & 271.3M & 13.2k & 276 & 9.04k \\ 

& Aragon & Ethereum & Service & ETH  & \cmark & \cmark & \cmark  &  \cmark & \cmark & \xmark & \xmark & \xmark & 199.1M & 14.2k & 606 & 1.03k \\ 

& Curve & Ethereum & Stablecoin & CRV  & \cmark & \cmark & \cmark  &  \cmark & \cmark & \xmark & \xmark & \xmark & 148.8M & 76.8k & 221 & 2.10k\\ 

& Fei & Ethereum & Stablecoin & TRIBE  & \cmark & \cmark & \cmark  &  \cmark & \cmark & \xmark & \xmark  & \xmark & 145.8M & 14.3k & 161 & 15.1k \\ 

& Decentraland & Polygon & NFTs & MANA  & \cmark & \cmark & \cmark  &  \cmark & \cmark & n/a & \xmark & \xmark & 138.5M & 308.9k & 2k & 94.7k \\ 

& Radicle & Ethereum & Service & RAD  & \cmark & \cmark & \cmark  &  \cmark & \cmark & \xmark & \xmark  & \xmark & 126.4M & 6.6k & 26 & 686 \\ 

& Aave & Polygon & DeFi (Lending) & AAVE  & \cmark & \cmark & \cmark  &  \cmark & \cmark & n/a & \xmark  & \cmark & 124.9M & 155.8k & 268 & 527.1k \\ 

& Compound & Ethereum & DeFi (Lending) & COMP  & \cmark & \cmark & \cmark  &  \cmark & \cmark & \xmark & \xmark  & \cmark & 121.5M & 208.6k & 169 & 13.4k \\ 

& DXdao & Polygon & DeFi (DEX) & DXD  & \cmark & \cmark & \cmark  &  \cmark & \cmark & n/a & \xmark  & \cmark & 117.1M & 1.4k & 915 & 2.54k \\ 

& Ribbon & Ethereum & DeFi (Derivative)  & RBN  & \cmark & \cmark & \cmark  &  \cmark & \cmark & \xmark &   \xmark  & \cmark & 116.3M & 4.4k & 31 & 4.75k  \\ 

& Synthetix & Ethereum & DeFi (DEX) & SNX  & \cmark & \cmark & \cmark  &  \cmark &\cmark  & \xmark &   \xmark  & \cmark & 115.3M & 91.5k & 569 &  14.6k\\ 

& MangoDAO & Solana & DeFi (DEX) & MNGO  & \cmark & \cmark & \cmark  &  \cmark & \cmark & n/a &   \xmark  & \cmark & 102.9M & 36k & 401 &  3.83k\\

& Gitcoin & Ethereum & Social network & GTC  & \cmark & \cmark & \cmark  &  \cmark & \cmark & \xmark &   \xmark  & \xmark & 92.2M & 33.7k & 144 & 70.4k \\ 

& Phala & Substrate & Polka's testnet & PHA  & \cmark & \cmark & \cmark  &  \cmark & \cmark & n/a & \xmark  & \xmark & 77.6M & 3.1k & 24 & 72\\ 

& Vesta.Fin & Polygon & Stablecoin & VSTA  & \cmark & \cmark & \cmark  &  \cmark & \cmark & n/a &   \xmark  & \xmark & 67.4M & 256.5k & 8 & 34.7k \\ 

& JPEG'd & Ethereum & DeFi (Lending) & JPEG  & \cmark & \cmark & \cmark  &  \cmark & \cmark & \xmark &   \xmark  & \cmark & 66M & 5.3k & 59 & 2.51k \\

& Euler.Fin & Ethereum & DeFi (Lending)  & EUL  & \cmark & \cmark & \cmark  &  \cmark & \cmark & \xmark &   \xmark  & \cmark & 63.5M & 2.6k & 55 &8.27k \\ 

& Merit Circle & Solana & NFTs & MC  & \cmark & \cmark & \cmark  &  \cmark & \cmark & n/a & \xmark  & \xmark & 61M & 8.9k & 26 & 2.83k \\ 

& SuperRare & Ethereum & NFTs & RARE  & \cmark & \cmark & \cmark  &  \cmark & \cmark & \xmark & \xmark  & \xmark & 54.1M & 8.7k & 17 & 1.23k \\ 

& KeeperDAO & Ethereum & DeFi (MEV-extractor) & ROOK  & \cmark & \cmark & \cmark  & \cmark & \cmark & \xmark &   \xmark  & \cmark & 53.5M & 17k & 41 & 1.21k  \\ 

& MakerDAO & Ethereum & Stablecoin & MKR  & \cmark & \cmark & \cmark  &  \cmark & \cmark & \xmark &   \xmark  & \xmark & 49.1M & 90.9k & n/a &  n/a \\ 

& UXDProtocol & Solana & Stablecoin & UXP  & \cmark & \cmark & \cmark  &  \cmark & \cmark & n/a &   \xmark  & \xmark & 49.6M & 11.7k & 819 & 3.27k  \\ 

& Yearn & Ethereum & DeFi (Lending) & YFI  & \cmark & \cmark & \cmark  &  \cmark & \cmark &  \xmark &   \xmark  & \cmark & 37.9M & 54.2k & 16 & 4.84k \\ 

& Balancer & Ethereum & DeFi (DEX) & BAL  & \cmark & \cmark & \cmark  &  \cmark & \cmark & \xmark &   \xmark  & \cmark & 36.1M & 45k & 378 & 82.1k \\ 

& PleasrDAO & Ethereum & NFTs & USDC  & \cmark & \cmark & \cmark  &  \cmark & \cmark & \xmark &   \xmark  & \xmark & 31.4M & 149 & 54 & 1.02k \\ 

& Sushiswap & Ethereum & DeFi (DEX) & SUSHI  & \cmark & \cmark & \cmark  &  \cmark & \cmark & \xmark &  \xmark  & \cmark & 28.7M & 109.1k & 290 &  49.2k\\ 

& Pangolin & Polygon & DeFi (DEX) & PNG  & \cmark & \cmark & \cmark  &  \cmark & \cmark & n/a &   \xmark  & \cmark & 19.2M & 32.5k & 45 &  2k\\ 

& 1inch & Ethereum & DeFi (DEX) & 1INCH  & \cmark & \cmark & \cmark  &  \cmark & \cmark & \xmark &   \xmark  & \cmark & 18.2M & 87.5k & 22 & 2.45k \\ 

& Lucidao & Polygon & Service & USDT  & \cmark & \cmark & \cmark  &  \cmark & \cmark &  n/a & \xmark  & \xmark & 11.8M & 1.3k & 6 &  154\\ 

& Kusama & Ethereum & Polka's testnet & KSM  &  \cmark  &  \cmark & \cmark  &  \cmark & \cmark & \xmark &   \xmark  &  \cmark  & 11.5M & 291.3k  & 863 & 5.65k \\ 

& Serum & Solana & DeFi (DEX) & SRM  &  \cmark &  \cmark & \cmark  &  \cmark & \cmark & n/a & \xmark  & \cmark & 4.5M & 226.1k & 52 & 246 \\  

& Bifrost & Substrate & DeFi (Lending) & BNC  &  \cmark &  \cmark & \cmark  &  \cmark & \cmark & n/a &   \xmark  & \cmark & 4M & 84.7k & 686 &  1.53k\\

\midrule

& \multicolumn{1}{c}{\rotatebox{0}{\textit{\textbf{Projects}}}} &  \multicolumn{2}{c}{\rotatebox{0}{\textit{\textbf{Field}}}}  & \multicolumn{2}{c}{\rotatebox{0}{\textit{\textbf{Note}}}}   
& \multicolumn{1}{c}{} & 
\multicolumn{4}{c}{\rotatebox{0}{\textit{\textbf{Projects}}}} & \multicolumn{3}{c}{\rotatebox{0}{\textit{\textbf{Field/Coverage}}}}   & \multicolumn{3}{c}{\rotatebox{0}{\textit{\textbf{Note}}}}\\

\midrule

\multirow{11}{*}{ \rotatebox{90}{\textit{\textbf{Tools \textcolor{gray}{\&} Launchpad}}}} & Aragon & \multicolumn{2}{c}{Management tools}  &   \multicolumn{2}{c|}{\multirow{9}{*}{\rotatebox{90}{\makecell{\textit{\textbf{Keywords}} on Guide, NFT, Arts, \\ Treasury,  Web3, DeFi, Game, \\ Analytics, DiD, Legal, Launcher, \\ Media,  Governance, Reputation,\\ 
Infrastructure, Social, Dispute }}}}  & \multicolumn{1}{c}{\multirow{11}{*}{\rotatebox{90}{\textit{\textbf{\textcolor{gray}{Related} DataFeeds}}}}}  &
\multicolumn{4}{c|}{Dune}   & \multicolumn{3}{c}{Data analytic}  &   \multicolumn{3}{c}{\url{https://dune.com/home}}    \\

& DAOStack &  \multicolumn{2}{c}{Management tools}  &   \multicolumn{2}{c|}{}  & \multicolumn{1}{c}{}  &
\multicolumn{4}{c|}{GraphQL}   & \multicolumn{3}{c}{Data analytic}  &   \multicolumn{3}{c}{\url{https://daostack.io}}    \\

& Colony&  \multicolumn{2}{c}{Management tools}  &   \multicolumn{2}{c|}{}  & \multicolumn{1}{c}{}  &
\multicolumn{4}{c|}{Colony API}   & \multicolumn{3}{c}{Data analytic}  &   \multicolumn{3}{c}{\url{https://colony.io}}    \\

& Snapshot &  \multicolumn{2}{c}{Off-chain voting platform}  &   \multicolumn{2}{c|}{}  & \multicolumn{1}{c}{}  &
\multicolumn{4}{c|}{DexTools}   & \multicolumn{3}{c}{Trading pair}  &   \multicolumn{3}{c}{\url{https://www.dextools.io}}   \\

& Tally &  \multicolumn{2}{c}{On-chain voting platform}  &   \multicolumn{2}{c|}{}  & \multicolumn{1}{c}{}  &
\multicolumn{4}{c|}{DefiLlama}   & \multicolumn{3}{c}{DeFi TVL aggregator}  &   \multicolumn{3}{c}{\url{https://defillama.com}}   \\

& DeepDAO &  \multicolumn{2}{c}{Information/aggregator}  &   \multicolumn{2}{c|}{}  & \multicolumn{1}{c}{}  &
\multicolumn{4}{c|}{TokenTerminal}   & \multicolumn{3}{c}{Projects, Financial data}  &   \multicolumn{3}{c}{\url{https://tokenterminal.com}}    \\

& DAOMasters &  \multicolumn{2}{c}{Launcher/Management}  &   \multicolumn{2}{c|}{}  & \multicolumn{1}{c}{}  &
\multicolumn{4}{c|}{RootData}   & \multicolumn{3}{c}{Fundraising, Investors}  &   \multicolumn{3}{c}{https://www.rootdata.com}    \\

& DAOlist &  \multicolumn{2}{c}{Information/aggregator}  &   \multicolumn{2}{c|}{}  & \multicolumn{1}{c}{}  &
\multicolumn{4}{c|}{CoinMarketCap}   & \multicolumn{3}{c}{Projects, Ranking}  &   \multicolumn{3}{c}{\url{https://coinmarketcap.com}}    \\

& Mirror &  \multicolumn{2}{c}{Publishing/Writing}  &   \multicolumn{2}{c|}{}  & \multicolumn{1}{c}{}  &
\multicolumn{4}{c|}{Zapper}   & \multicolumn{3}{c}{DAOs, NFTs, DeFi}  &   \multicolumn{3}{c}{\url{https://zapper.xyz/daos}}    \\

& Gnosis Safe &  \multicolumn{2}{c}{Multisig wallets}  &   \multicolumn{2}{c|}{}  & \multicolumn{1}{c}{}  &
\multicolumn{4}{c|}{DappRadar}   & \multicolumn{3}{c}{DApps, NFTs, DeFi}  &   \multicolumn{3}{c}{\url{https://dappradar.com}}    \\

& IPFS &  \multicolumn{2}{c}{Storage infrastructure}  &   \multicolumn{2}{c|}{}  & \multicolumn{1}{c}{}  &
\multicolumn{4}{c|}{DexScreener}   & \multicolumn{3}{c}{Trading pair, Price}  &   \multicolumn{3}{c}{\url{https://dexscreener.com}}    \\

\bottomrule
\end{tabular}
  }
  \begin{tablenotes}
       \footnotesize
       \item[] \quad\quad \textcolor{magenta}{\ding{43}} Source data in this paper mainly refers to DeepDAO (\url{https://deepdao.io/organizations}) [May 2023].
       \item[] \quad\quad \textcolor{magenta}{\ding{43}} \textcolor{black}{\textbf{Insights}}-\ding{207} Ethereum DAOs (post-Merge) are censored due to the OFAC-compliant blocks (MEV Watch \url{https://www.mevwatch.info}). 
       \item[] \quad\quad\quad\quad\quad\quad\quad   \ding{208} DeFi-related DAOs are incentive-compatible as stakeholders are motivated to hold and use tokens to maximize their profits. 
     \end{tablenotes}
  \end{threeparttable}  

\end{table*}

\end{document}